\documentclass[%
    reprint,
    superscriptaddress,
    aps,
    pre,
    preprintnumbers,
]{revtex4-2}
\usepackage{amsmath}
\usepackage{amsfonts}
\usepackage{empheq}
\usepackage{hyperref}
\usepackage[protrusion=true,expansion=false]{microtype}
\usepackage{float}
\usepackage{graphicx}
\usepackage{caption}
\usepackage{subcaption}
\usepackage{natbib}
\usepackage{array}
\usepackage{standalone}
\usepackage{tikz}

\begin{document}

\title{
    Complex surface patterning in homo- and heteroepitaxial contexts:\\
    (simultaneous) step bunching and step meandering
}

\author{Vassil Ivanov}
\email[Corresponding author: ]{vvasilevi@fmi.uni-sofia.bg}
\affiliation{Faculty of Mathematics and Informatics, Sofia University, 1164
    Sofia, Bulgaria}

\author{Vesselin Tonchev}
\affiliation{Faculty of Physics, Sofia University, 1164 Sofia, Bulgaria}

\author{Marta A. Chabowska}
\affiliation{Institute of Physics, Polish Academy of Sciences, al. Lotnik\'{o}w
    32/46, Warsaw, Poland}

\author{Hristina Popova}
\affiliation{Institute of Physical Chemistry, Bulgarian Academy of Sciences,
    Acad. G. Bonchev str., 1113 Sofia, Bulgaria}

\author{Magdalena A. Za{\l}uska-Kotur}
\affiliation{Institute of Physics, Polish Academy of Sciences, al. Lotnik\'{o}w
    32/46, Warsaw, Poland}

\date{\today}

\begin{abstract}
    We confront a meso-scale continuum model, archetypical for the
    heteroepitaxial context, with an atomistic Vicinal Cellular Automaton
    (VicCA), built as a homoepitaxial counterpart, to show that in (2+1)D
    complex surface instabilities are fundamental growth phenomena rather than
    context-specific artifacts. Our approach is to first construct a
    Ginzburg-Landau-type model, designed to extend the previously (1+1)D
    Tersoff-type models to (2+1)D. We complement the continuum approach with a
    discrete one - the VicCA, in which we use a novel version of the potential
    landscape for the diffusing particles - a double-well potential located at
    each step edge. Importantly, this framework also captures both step bunching
    and step meandering. While these behaviors are rarely treated together in
    earlier approaches, they naturally coexist in real materials. Thus we
    establish a cross-context correspondence at the level of obtained
    morphologies and morphology diagrams and, additionally, a multiscale
    perspective on the governing parameters, bridging the gap between the
    mesoscale and atomistic modeling.
\end{abstract}


\maketitle

\section{Introduction}

The coexistence of step bunching and step meandering poses a long-standing
problem in the physics of unstable vicinal crystal surfaces \citep{deTheije2000,
ohtani2000step, neel2003meandering, omi2005new, Yu2011, Krzyewski2014, Huo2024}.
On the other hand, the regular step-flow growth is important from a practical
point of view in device manufacturing. It allows for precise surface morphology
control and the potential to avoid the stochastic nature of island growth
\citep{kangawa2026chemical}.

This regular step-flow growth occurs in two principally different experimental
realizations - the technologically relevant layers are grown on a substrate that
is either the same or chemically different from the growing layers. The success
story of the blue diode, starting from depositing the first smooth layer of
AlN/sapphire is well-known and presented elsewhere
\citep{Amano2014NobelLecture}. For the purposes of the present work we should
only note that although this is a story about heteroepitaxy, the fine details
reveal attempts to decrease the misfit in this heteroepitaxial system. In the
recent experimental works of \citet{Huo2024}, that focus on the heteroepitaxial
growth of GaN on SiC substrate, the authors have demonstrated that in this
context both bunching and meandering are possible. Unfortunately, on the
theoretical side, the simultaneous observation of step bunching and meandering
in \textit{heteroepitaxial} systems is not reported. Still, there is a
well-understood (1+1)D framework to study step bunching only due to Tersoff et
al.~\citep{Tersoff1995, Tersoff1997}. It is important to note that in this model
there are two types of step-step interaction - logarithmically increasing
attraction (due to the adlayer) and the canonical elastic step-step repulsion.
Thus the problem is how to go consistently beyond, into (2+1)D, with minimal
additional prerequisites.

On the homoepitaxial side, both experimental and modeling studies provide better
and deeper perspectives on \textit{formally} the same phenomenon. There are
several experimental studies focusing on SiC, Cu-vicinals, GaAs, and Si(111)
\citep{ohtani2000step, neel2003meandering, omi2005new, Galiana2013}. In some
cases, step bunching and step meandering are present in the experimental results
\citep{deTheije2000}, but are not explicitly named and meanders are treated as
an artifact. These studies highlight the opposite roles of the direct and
inverted Ehrlich-Schwoebel effect in the homoepitaxial context - where one
destabilizes, the other stabilizes, and vice versa. The modeling community
responded to this challenge with the works of \citet{LIN2007, Yu2011,
Krzyewski2014}.

As a matter of definitions that we will use in the present text, step bunching
is the instability where the steps group together, forming bunches with spacings
less than the initial vicinal distance, separated by wide terraces. Step
meandering is the growth mode where the straight step edge loses its stability
and obtains a curvilinear shape, leading to complex interactions and a
quasi-stochastic surface morphology. These two phenomena can be thought of as
complementary ones - proceeding in two orthogonal directions: step bunching
along the direction of step flow and step meandering across the direction of
step flow (normal and tangential to the step edge directions, respectively). It
is the complementarity of the two mechanisms that permits them to occur
separately or in a concerted fashion \citep{Misbah2010}.

These instabilities may require modifications of the growth strategies, thus
developing adequate models to build surface morphology diagrams becomes crucial
for technological success. Prior studies focus primarily on the step bunching
mode, with the most significant models of step dynamics \citep{Liu1998,
Tersoff1995, Krasteva2016} treating the system as points on a line that attract
and repel each other with forces proportional to (often non-linear) functions of
terrace widths (step-to-step distances). Coarse-grained models
\citep{Stoyanov1998, Stoyanov1998b, PTVV2002, Krug2005} yield smooth versions of
the overall surface profile and, more crucially, permit scaling analysis, thus
creating a quantitative framework of universality classes against which all past
and future models should be assessed. These approaches represent vicinal
surfaces as a (1+1)D staircase-like structure.

In more recent times, the so-called VicCA models have been developed, which
approach the problem of homoepitaxial growth from an atomistic point of view
\citep{zaluska-kotur_step_2021, Chabowska-JCG, Chabowska-PRB,
chabowska_surface_2025, Redkov2025}. VicCA models are a combination of a
Cellular Automaton (CA), which implements the basic terrace-ledge-kink growth
rules, complemented by a Monte Carlo module to simulate the diffusion of adatoms
through the disordered phase. These models have been successful in obtaining
quantitative estimates of the growth regimes in terms of scaling laws. The main
weakness of VicCA models is that their results are only available through
numerical investigation, and it is often intractable to derive quantitative
criteria to determine the surface (in-)stability. Nonetheless, their flexibility
is such that we put a special focus on VicCA models in the present work.

We follow two distinct pathways through the concepts of surface instabilities
and their combinations. First, we develop a continuum model based on the
approach of

\citet{Kandel1994},~\citep{Kandel1994-2}, in which the destabilization
``engine'' originally describes impurity-driven step bunching. We replace this
engine with a minimal model \citep{Krasteva2016} from a class of models
describing unstable step motion in heteroepitaxy \citep{Tersoff1995,
Krasteva2016}. We then investigate the resulting model both analytically,
following the approach of \citet{Connell2004}, and numerically. The proposed
model contains a minimal set of parameters - step stiffness, the magnitudes and
ranges of attraction and repulsion, while having a convenient structure that
allows for the simulation of large systems. The nearest-neighbor interactions in
the engine of destabilization yield, when locally linearized, tridiagonal
matrices that enable efficient computations.

The step interaction function \({h(w) = \kappa_r w^{-q} - \kappa_a w^{-p}}\),
with \(p < q\), is a force of Lennard-Jones type \citep{Luo2016-gd, Luo2021}
where attraction dominates at larger distances and the repulsion is
shorter-ranged. Here, \(w\) is the terrace width, \(\kappa_a\) and \(\kappa_r\)
set the attraction and repulsion magnitudes, respectively, and \(p,q > 0\) are
the corresponding interaction exponents. The step stiffness or, effectively, the
line-tension of a step, opposes its bending. We show then that the coexistence
of bunching and meandering requires the system to be driven beyond a natural
spacing \(w_M\) (provided that \(p < q\)) after which both types of structures
develop from the same underlying potential and are different manifestations of
the coarsening behavior. For convenience, we divide the parameter space into
regions that exhibit bunching, meandering, their coexistence, and regular
patterns as well. Then, our efficient GPGPU-based approach of solving the
differential-difference equations permits us to analyze the long-time evolution
of the surfaces studied, and while the present work focuses on the model
proposition, qualitative and computational aspects of the analysis, the rigorous
proofs of the gradient-flow structure, equilibrium classification, meandering
nature and coarsening rates are developed in a companion work
\citep{Ivanov2026}.

Further, we confront these findings with a sophisticated \textit{ad hoc} version
of our VicCA framework, aimed at homoepitaxial vicinal surface instabilities
\citep{zaluska-kotur_step_2021, Chabowska-PRB, chabowska_surface_2025}. We
introduce a modified potential energy landscape - a double-well one, located at
each evolving step, that shapes the diffusion field. This complex environment
resulting from the coupling of morphology and adatom distribution gives rise to
little-studied surface patterns. Notably, the morphology diagram reveals regions
that closely resemble those found in the continuum-model description.

\section{The (2+1)D continuum step model}

We shall follow the overall reasoning provided by K\&W in \citep{LWJeong1997} to
obtain the general form of the model of interest. Let
\(\{u_n(t,x)\}_{n=0}^{N-1}\) be the positions of \(N\) steps at time \(t\),
where \(x\) is the coordinate along the step edge (perpendicular to the step
flow direction). The steps are ordered such that \(u_n < u_{n+1}\) for all \(x\)
at \(t = 0\). The system Hamiltonian is then given by:
\begin{equation}
    H \left( \{u_n\} \right) = \int_0^{L} \sum_{n=0}^{N-1} \left[ \frac{\tilde \beta}{2} \left( \frac{\partial u_n}{\partial x} \right)^2 + V (\Delta u_n) \right] dx \label{eq:main_hamiltonian}
\end{equation}
where \(\tilde \beta\) is the step stiffness, \(V\) is the interaction potential
between steps \(n - 1\) and \(n\), and \(\Delta u_n = u_n - u_{n-1}\) is the
terrace width. Then, we can define the system's chemical potential as:
\begin{equation}
    \mu_n(x) = \Omega \left[ V'(\Delta u_{n+1}) - V'(\Delta u_n) + \tilde{\beta} \frac{\partial^2 u_n}{\partial x^2} \right]
\end{equation}
where \(\Omega\) is the atomic area. Following K\&W, we will focus on the case
where the steps' interaction is through a reservoir of constant chemical
potential, i.e.:
\begin{equation}
    \frac{d u_n}{d t} \propto \mu_n(x) - \mu_{res}
\end{equation}
where \(\mu_{res}\) is the chemical potential of the reservoir. This leads to
the following general partial differential equation (PDE) system for the step
positions \citep{LWJeong1997}:
\begin{equation}
    \label{eq:general_pde_form}
    \frac{\partial u_n}{\partial t}
    = \gamma \frac{\partial^2 u_n}{\partial x^2}
      + f\left( u_{n-s},\ldots, u_{n},\ldots, u_{n+r} \right), \quad n
    = 0, \ldots, N - 1
\end{equation}
where \(\gamma\) is a constant proportional to the step stiffness, and \(f\) is
a generalized step-velocity function, which is, in general, not symmetric in its
arguments and \(s \ne r\). Note that, in general, \(\gamma\) can also be
anisotropic and depends on the miscut angle \(\theta\), temperature, etc. Here
the works of \citep{Stasevich2007, Margetis2008, Yu2011, Krukowski2022} will
serve as a basis for future extensions of the model.

\autoref{eq:general_pde_form} directly incorporates the two ``orthogonal''
contributions - the step-velocity models the step-step attraction/repulsion
effects that lead to step bunching and the second derivative term which models
the stabilizing effect of step-stiffness that resists transverse fluctuations.

Selecting a specific form of \(f\) leads to different models with different
properties, and thus we will focus on a minimal model that can reproduce the
main features of heteroepitaxial step bunching, the so-called \textit{MM2} model
due to Tonchev~\citep{Krasteva2016}. In this way we obtain the following
specific form of the model:
\begin{equation}
    \label{eq:mm2_meandering}
    \frac{\partial u_{n}}{\partial t}
    = \gamma \frac{\partial^{2} u_{n}}{\partial x^{2}}
      - \kappa_{a}\left( \Delta u_{n}^{-p} - \Delta u_{n+1}^{-p} \right)
      + \kappa_{r}\left( \Delta u_{n}^{-q} - \Delta u_{n+1}^{-q} \right)
\end{equation}
where \(\kappa_a\) and \(\kappa_r\) are constants proportional to the strength
of the step-step attraction and repulsion respectively, and \(p, q\) are
positive exponents that determine the distance dependence of the step-step
interactions. As a matter of convention, we take the surface height to increase
with the step index \(n\), and define the forward direction as increasing \(u\).
Then, for the \(n\)-th step with position \(u_n\), \(\Delta u_n\) is the width
of the lower terrace and \(\Delta u_{n+1}\) is the width of the upper terrace.

\autoref{eq:mm2_meandering} is subject to the initial condition
\(u_n(0,x) = l_0 n\), where \(l_0\) is the initial vicinal distance, periodic
boundary condition along \(x\) and shifted periodic boundaries along \(n\):
\(u_{n+N} = u_{n} + N l_0\). This results in an \textit{ad hoc} model for step
bunching and meandering which will be the main focus of study in the present
work.

Before we proceed with analyzing the mathematical structure of this newly
proposed model, we will first make a few notes about the physical context it is
applicable for - it comes from step bunching on strained heteroepitaxial
systems. When a layer grows on a substrate with a different lattice parameter,
the lattice mismatch produces elastic stress. As shown by \citet{Tersoff1995},
elastic relaxation at the steps generates an attractive interaction that is
longer-ranged than the omnipresent step-step repulsion and can therefore drive
step bunching. Strain-driven bunching has also been studied experimentally in
heteroepitaxial systems \citep{Zhu1999, Yoon2007} and these works correspond
well with the Tersoff model. This model also exhibits the so-called B1-type
bunching, in which the minimum step-to-step distance remains approximately
independent of the number of steps in the bunch \citep{Tonchev2012}, i.e. the
bunches are incompressible. The developed bunches retain an approximately
constant interior slope as they grow and meet the terraces at a finite angle.

For \(p = 1\) and \(q = 3\), the interaction terms in \textit{MM2} correspond to
a nearest-neighbour truncation of the elastic step-step force in the Tersoff
model. The attraction therefore represents the effect of misfit stress, with its
relative magnitude \(\kappa_a\). These interactions are then combined with local
relaxation kinetics and step stiffness to obtain a minimal description of
bunching and meandering in this heteroepitaxial setting - our (2+1)D model in
\autoref{eq:mm2_meandering}. The corresponding interaction potential is
discussed in greater detail in \autoref{sect:structure}.

\textit{MM2} also exhibits B1-type bunching and offers a practical advantage for
numerical studies: for the straight-step models with the interaction exponents
used here, the time scaling law of the number of steps in the bunch for
\textit{MM2} is \(N_b \sim t^{1/3}\), while for
\citet{Tersoff1995},~\citep{Krasteva2016} it is \(N_b \sim t^{1/5}\). Here
\(N_b\) is the number of steps in a bunch. The simpler dynamics and faster
coarsening make \textit{MM2} convenient for extending the present numerical
study to meandering steps. Finally, as it will become clear, this simplified
model presents additional mathematical structure, that is not available in the
more general Tersoff model and thus allows us to gain deeper insights on these
instabilities in the heteroepitaxial context.

\section{Structural properties}
\label{sect:structure}

We now proceed by shedding light on some of the mathematical and structural
properties of our continuum model. It is directly clear that it has a
gradient-flow structure as shown by \autoref{eq:main_hamiltonian}. In
non-dimensional variables (see \citet{Ivanov2026} for details) the system can be
written as \({\partial_t u_n = - \delta \mathcal F/\delta u_n}\) where
\(\mathcal{F}\) is the Ginzburg-Landau type functional:
\begin{equation}
    \label{eq:ginzburg-landau}
    \mathcal{F}[u] = \sum_{n=0}^{N-1} \int_0^L \left\{ \frac{\varepsilon}{2}\left( \frac{\partial u_n}{\partial x} \right)^2 + \varepsilon^{-1} V(\Delta u_n) \right\} dx
\end{equation}
Here, \(V'(w) = - h(w)\) with \(h(w) = w^{-q} - w^{-p}\) \citep{Luo2016-gd}. In
the case when the exponents are \(p = 1\) and \(q = 3\), which we use below for
numerical results, this corresponds to logarithmic attraction in the potential
\(V(w)\) and an inverse-square repulsive potential. \citet{Tersoff1995} have
associated the logarithmic term with treating the steps as ``force monopoles'',
while the inverse-square term is the canonical elastic repulsion between steps.

Then, due to this gradient-flow nature, the energy monotonically decreases along
the trajectories:
\begin{equation}
    \frac{d \mathcal F}{d t} = - \sum_{n=0}^{N-1} \int_0^L \left| \partial_t u_n \right|^2 dx
                             \le 0
\end{equation}

From monotonicity it follows that in this model macrosteps
(\(\Delta u_n \rightarrow 0\)) cannot be formed, since
\({V(w \rightarrow 0^+) = + \infty}\). Thus, any trajectory approaching
\(\Delta u_n = 0\) would require infinite energy, contradicting the monotonicity
condition above. This is consistent with the so-called ``step-step exclusion''
principle from which important results, based on the fermionic treatment of
steps, have been derived by \citet{Akutsu1988-fe, Joos1991-iv}. This also
corresponds with the \(\mathcal O(1)\) scaling of \(l_{min}\) obtained by
\citet{Luo2016-gd, Luo2021} and confirms that the B1-type of the model is
preserved when extending the approach from 1D to 2D.

Another important observation that we make about \autoref{eq:mm2_meandering} is
that the model redistributes the average distance between steps, rather than
having a net ``advection'' of the average step position. This can be readily
seen by summing over \(n\), noticing that the sum on the right-hand side
telescopes, leaving:
\begin{equation}
    \partial_t \bar u = \gamma \partial_{xx} \bar u, \qquad \bar u
                      = \frac{1}{N} \sum_{n} u_n
\end{equation}
This simple conservation of total length provides a useful numerical diagnostic
as well: deviations from pure diffusion of the average \(\bar u\) indicate
solver error rather than true physical coarsening.

\begin{figure}
    \centering
    \includegraphics[width=0.9\linewidth]{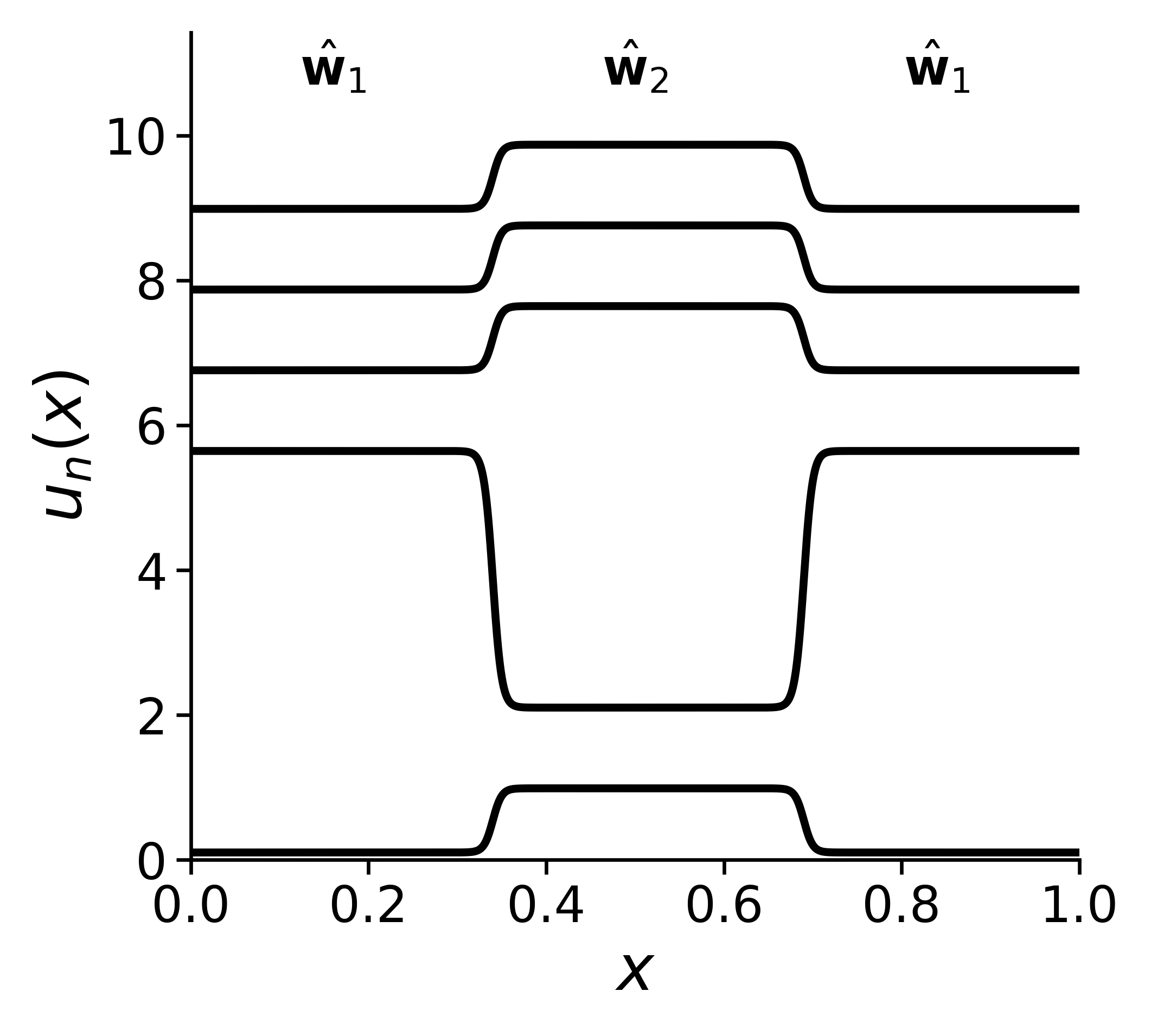}
    \caption{
        Illustration of the domain split into regions corresponding to different
        elements of \(\mathcal E_1^5\) connected with thin interfaces.
    }
    \label{fig:domain_decomposition}
\end{figure}

We will now note some of the deeper properties of the model proved in detail in
\citet{Ivanov2026}, which are nonetheless relevant to the structure of the
numerical results presented below. We will follow on a high level the
matched-asymptotics approaches of \citet{Rubinstein1989, Connell2004}: we start
by asymptotically expanding the full PDE in powers of \(\varepsilon\). Then, at
\(\varepsilon = 0\) we will obtain the original ODE \textit{MM2} model as a
zeroth order term in the expansion. This point can be illustrated by first
defining the stretched coordinate \(s \coloneqq t/\varepsilon\) to obtain:
\begin{equation}
    \partial_s u_n = \varepsilon^2\partial_{xx}u_n
                     - \left[ h(w_{n+1}) - h(w_n) \right]
\end{equation}
Then, we introduce the outer expansion:
\({u_n = u_n^0 + \varepsilon^2 u_n^1 + \mathcal O (\varepsilon^4)}\) and at
zeroth and first order we have respectively:
\begin{equation}
    \label{eq:mm2_ode_nondimensionalized}
    \partial_s u_n^{(0)} = - \left[ h(w_{n+1}^{(0)}) - h(w_n^{(0)}) \right]
\end{equation}
\begin{equation}
    \partial_s u_n^{(1)}
    = \partial_{xx}u_n^{(0)}
      - \left[ h'(w_{n+1}^{(0)})w_{n+1}^{(1)} - h'(w_n^{(0)})w_n^{(1)} \right]
\end{equation}
\autoref{eq:mm2_ode_nondimensionalized} is exactly the ODE \textit{MM2} model in
non-dimensionalized form. The central point of \citet{Rubinstein1989} is that
the coarsening dynamics as \(\varepsilon \rightarrow 0\) can be understood by
first enumerating the stable equilibria of the ODE model. For the full nonlinear
PDE system, different spatial regions (intervals along the \(x\)-axis) then
settle into different stable equilibria. These spatial regions are connected
with thin interfaces, where most of the ``excess energy'' lies (cf.
\autoref{fig:domain_decomposition} for an illustration).

Similarly to \citet{Connell2004}, to simplify the analysis, we focus on the
equilibria of the equivalent ODE for the terrace width evolution:
\begin{equation}
    \frac{d w_{n}}{d s} = -\left[ h(w_{n+1}) - 2 h(w_{n}) + h(w_{n-1}) \right]
\end{equation}
These equilibria are then given as solutions of the following discrete
Laplace-type equation under periodic boundaries \(w_{n+N} = w_n\):
\begin{equation}
    h(w_{n+1}) - 2 h(w_{n}) + h(w_{n-1}) = 0
\end{equation}
Solving this recurrence relation it can be shown that for a system of \(N\)
steps, the equilibria are \(N\)-dimensional points consisting of \(k\) ``long''
terraces of length \(w_L\) and \(N - k\) ``short'' terraces with length \(w_S\).
For example, for \(N = 5, k = 1\) the tuple \((w_L, w_S, w_S, w_S, w_S)\) and
all of its permutations are all valid equilibria of the system
\citep{Connell2004, Ivanov2026}. We annotate the set of such equilibria as
\(\mathcal E_k^N\).

Using the above constructions, it can be shown \citep{Ivanov2026} that, when
\(l_0 > w_M\), only equilibria from \(\mathcal E_1^N\) are stable, while all
other equilibria from \(\mathcal E_k^N\), for \(k > 1\), are saddle points of
the potential. Here \(w_M\) is the length at which \(h(w)\) attains its minimum.
Additionally, in the linear analysis below, it will be shown that \(l_0 > w_M\)
is also the instability condition for the uniformly spaced initial
configuration. \autoref{fig:equilibria_ode_mm2} schematically illustrates this
discrete structure.

\begin{figure}[t]
    \centering

    \begin{subfigure}{\columnwidth}
        \centering
        \scalebox{0.7}{%
\begin{tikzpicture}[
    level/.style={draw=black, line width=0.8pt},
    trans/.style={<->, black, line width=1.1pt},
    every node/.style={font=\large}
]

\begin{scope}[xshift=0cm]
    \foreach \y in {4,2,1,0,-1}
        \draw[level] (-1.4,\y) -- (1.4,\y);

    \draw[trans] (0,2.05) -- (0,3.95);
    \draw[trans] (0,1.05) -- (0,1.95);
    \draw[trans] (0,0.95) -- (0,0.05);
    \draw[trans] (0,-0.05) -- (0,-0.95);

    \node[anchor=west] at (0.30,3.00) {$w_{\mathrm{L}}$};
    \node[anchor=west] at (0.30,1.45) {$w_{\mathrm{S}}$};
    \node[anchor=west] at (0.30,0.45) {$w_{\mathrm{S}}$};
    \node[anchor=west] at (0.30,-0.55) {$w_{\mathrm{S}}$};
\end{scope}

\begin{scope}[xshift=4.2cm]
    \foreach \y in {4,3,1,0,-1}
        \draw[level] (-1.4,\y) -- (1.4,\y);

    \draw[trans] (0,3.05) -- (0,3.95);
    \draw[trans] (0,1.05) -- (0,2.95);
    \draw[trans] (0,0.05) -- (0,0.95);
    \draw[trans] (0,-0.05) -- (0,-0.95);

    \node[anchor=west] at (0.30,3.48) {$w_{\mathrm{S}}$};
    \node[anchor=west] at (0.30,1.90) {$w_{\mathrm{L}}$};
    \node[anchor=west] at (0.30,0.43) {$w_{\mathrm{S}}$};
    \node[anchor=west] at (0.30,-0.55) {$w_{\mathrm{S}}$};
\end{scope}

\end{tikzpicture}%
}
        \caption{\(\mathcal{E}_1^5\)}
        \label{fig:equilibria-e1-5}
    \end{subfigure}

    \vspace{0.2em}

    \begin{subfigure}{\columnwidth}
        \centering
        \scalebox{0.7}{%
\begin{tikzpicture}[
    level/.style={draw=black, line width=0.8pt},
    trans/.style={<->, black, line width=1.1pt},
    every node/.style={font=\large}
]

\begin{scope}[xshift=0cm]
    \foreach \y in {4,2.45,1,0,-1}
        \draw[level] (-1.4,\y) -- (1.4,\y);

    \draw[trans] (0,2.50) -- (0,3.95);
    \draw[trans] (0,1.05) -- (0,2.40);
    \draw[trans] (0,0.05) -- (0,0.95);
    \draw[trans] (0,-0.05) -- (0,-0.95);

    \node[anchor=west] at (0.30,3.05) {$w_{\mathrm{L}}$};
    \node[anchor=west] at (0.30,1.55) {$w_{\mathrm{L}}$};
    \node[anchor=west] at (0.30,0.43) {$w_{\mathrm{S}}$};
    \node[anchor=west] at (0.30,-0.55) {$w_{\mathrm{S}}$};
\end{scope}

\begin{scope}[xshift=4.2cm]
    \foreach \y in {4,3,1.4,0,-1}
        \draw[level] (-1.4,\y) -- (1.4,\y);

    \draw[trans] (0,3.05) -- (0,3.95);
    \draw[trans] (0,1.45) -- (0,2.95);
    \draw[trans] (0,1.35) -- (0,0.05);
    \draw[trans] (0,-0.05) -- (0,-0.95);

    \node[anchor=west] at (0.30,3.48) {$w_{\mathrm{S}}$};
    \node[anchor=west] at (0.30,2.10) {$w_{\mathrm{L}}$};
    \node[anchor=west] at (0.30,0.72) {$w_{\mathrm{L}}$};
    \node[anchor=west] at (0.30,-0.55) {$w_{\mathrm{S}}$};
\end{scope}

\end{tikzpicture}%
}
        \caption{\(\mathcal{E}_2^5\)}
        \label{fig:equilibria-e2-5}
    \end{subfigure}

    \caption{
        Schematic representation of the equilibrium states for a system with
        \(N = 5\). \autoref{fig:equilibria-e1-5} represents two valid equilibria
        from the \(\mathcal E_1^5\) set, while \autoref{fig:equilibria-e2-5}
        represents valid equilibria from \(\mathcal{E}_2^5\). Within each
        sub-figure, the two equilibrium states are equivalent up to a cyclic
        permutation. Note that due to the periodic boundaries the ``fifth''
        length is implied.
    }
    \label{fig:equilibria_ode_mm2}
\end{figure}

We can further observe that elements of the same set \(\mathcal E_k^N\) are
energetically degenerate, as:
\begin{equation}
    \sum_n V(\hat w_n) = k V(w_L)
                         + (N - k) V(w_S), \qquad \forall \mathbf{\hat w}
                       \in \mathcal{E}_k^N
\end{equation}
where \(\hat w_n \in \{w_S, w_L\}\) and \(\mathbf{\hat w}\) is a concrete
equilibrium \(N\)-dimensional point from \(\mathcal{E}_k^N\).

Now, we focus again on the (2+1)D PDE model. Suppose two regions in space have
settled into two distinct members of \(\mathcal E_1^N\) (cf.
\autoref{fig:domain_decomposition}), the energy jump between both sides of the
interface connecting them is zero:
\({[V] = V(\mathbf{\hat w_1}) - V(\mathbf{\hat w_2}) = 0}\), for
\(\mathbf{\hat w_1}, \mathbf{\hat w_2} \in \mathcal E_1^N\).

Through slow manifold theory for models of the kind of
\autoref{eq:ginzburg-landau}, developed in depth by \citet{Rubinstein1989,
Bronsard1990, Bethuel2011}, it can be shown that, for
\(\varepsilon \rightarrow 0\), these interfaces evolve exceedingly slowly -
slower than any algebraic power of \(\varepsilon\). From a numerical point of
view, once the system enters this slow manifold, the interfaces move exceedingly
slowly and the meandered profiles appear effectively frozen over any realistic
simulation time. Thus, although curved steps are ``penalized'' in the energy
functional and the true equilibrium is always straight, bunched steps, meanders
persist as long-lived metastable structures that appear stable in simulation.

In summary for our continuum model, equilibrium configurations depend on the
finite system and its boundary conditions. Our periodic cell fixes the total
terrace width to \(Nl_0\), which constrains the possible bunched states. Once
bunches or meander domains approach the cell size, finite-size effects matter.
However, despite this, slow motion analysis shows that meanders can persist for
long times because the interfaces between different close-to-equilibrium regions
evolve slowly. Therefore all numerically obtained patterns need to be
interpreted with that context.

\section{Linear stability analysis}

It is clear that the system given by \autoref{eq:mm2_meandering} has a trivial
solution:
\begin{equation}
    u_n(t,x) = l_0 n
\end{equation}
where \(l_0\) is the initial step-step spacing. Thus, an equally spaced
configuration is an equilibrium point for the system. To study the stability of
this solution, we shall introduce a small perturbation \(\xi_n(t,x)\) such that:
\begin{equation}
    u_n = n l_0 + \xi_n(t,x)
\end{equation}
\begin{equation}
    \Delta u_n = l_0 + \left( \xi_n - \xi_{n-1} \right) = l_0 + \delta_n(t,x)
\end{equation}
where $\left| \delta_n \right| \ll l_0$.

Now using the Taylor expansion:
\begin{equation}
    (l_0 + \delta )^{-\varphi} = l_0^{-\varphi} \left( 1 - \varphi \frac{\delta}{l_0} \right)
                                 + O(\delta^2)
\end{equation}
we can linearize \autoref{eq:mm2_meandering} about the trivial solution to
obtain an equation for the perturbation evolution:
\begin{equation}
    \frac{\partial \xi_n}{\partial t}
    = \gamma \frac{\partial^2 \xi_n}{\partial x^2}
      + \left( \kappa_a p l_0^{-\left( p+1 \right)} - \kappa_r q l_0^{-\left( q+1 \right)} \right) \left( \delta_n - \delta_{n+1} \right)
\end{equation}
Setting
\(c = \kappa_a p l_0^{-\left( p+1 \right)} - \kappa_r q l_0^{-\left( q+1 \right)}\)
and expanding $\delta_n, \delta_{n+1}$:
\begin{equation}
    \label{eq:linearized_perturbation_eq}
    \frac{\partial \xi_n}{\partial t}
    = \gamma \frac{\partial^2 \xi_n}{\partial x^2}
      - c \left( \xi_{n+1} - 2 \xi_n + \xi_{n-1} \right)
\end{equation}
\autoref{eq:linearized_perturbation_eq} shows a surprisingly simple structure -
the step-step interaction terms lead to a discrete Laplacian in the step index
\(n\), while the step curvature term leads to a continuous Laplacian in the
coordinate along the step edge \(x\). This allows us to rewrite the system in
matrix form. Let \({\bar{\xi} = (\xi_0, \xi_1, \xi_2, \ldots)}\):
\begin{equation}
    \frac{\partial \bar{\xi}}{\partial t} = \gamma\frac{\partial^2 \bar{\xi}}{\partial x^2}
                                            + A \bar \xi
\end{equation}
where \(A = -c T\) and \(T\) is the second-difference matrix
\citep{higham2022seconddiff}, thus the component equations become:
\begin{equation}
    \label{eq:component_eq}
    \partial_t \xi_n = \gamma \partial_{xx} \xi_n + (A\bar\xi)_n, \quad n
                     = 0,\ldots,N - 1, x
                     \in[0,L]
\end{equation}
where periodic boundaries along \(x\) are imposed. Now we assume the
Fourier-type solution for each component $\xi_n$:
\begin{equation}
    \xi_n = \sum_{j,m} \alpha_{j,m} (t) e^{i k_j x} e^{2 \pi i \frac{m}{N} n}
\end{equation}
With \(k_j = 2 \pi j / L\), \(j \in \mathbb{Z}\) and \(m = 0, \ldots, N - 1\),
plugging this for each of the right-hand-side components of
\autoref{eq:component_eq} we obtain:
\begin{equation}
    \partial_{xx} e^{i k_j x} = - k_j^2 e^{i k_j x}
\end{equation}
\begin{align}
    (A\bar{\xi})_n & = -c \left( e^{\frac{2\pi i m}{N}} + e^{-\frac{2\pi i m}{N}} - 2 \right)\xi_n \notag \\
                   & = 4c \sin^2\!\left( \frac{\pi m}{N} \right)\xi_n = \lambda_m \xi_n
\end{align}
for \(m = 0, \ldots, N - 1\). After substituting the full expansion into
\autoref{eq:component_eq}, we obtain:
\begin{equation}
    \sum_{j,m} \frac{d\alpha_{j,m}}{dt} e^{i k_j x} e^{\frac{2\pi i m}{N} n} = \sum_{j,m} \left( \lambda_{m} - \gamma k_j^{2} \right) \alpha_{j,m}(t) e^{i k_j x} e^{\frac{2\pi i m}{N} n}
\end{equation}
After solving the initial value problem (IVP) for the \(\alpha_{j,m}\)
coefficients, we obtain:
\begin{equation}
    \label{eq:fourier_expansions}
    \xi_n(t,x) = \sum_{j \in \mathbb{Z}} \sum_{m=0}^{N-1} \alpha_{j,m}(0)\, e^{(\lambda_m-\gamma k_j^{2})t}\, e^{i k_j x}\, e^{\frac{2\pi i m}{N} n}
\end{equation}
The \(\alpha_{j,m}(0)\) coefficients are obtained from the initial conditions by
projecting them in the Fourier basis:
\begin{equation}
    \alpha_{j,m}(0) = \frac{1}{N L} \sum_{n=0}^{N-1} \int_{0}^{L} \xi_n(0, x)\, e^{-i k_j x}\, e^{-\frac{2\pi i m}{N} n} dx
\end{equation}
The amplitude of each Fourier mode grows/decays exponentially with rate:
\begin{equation}
    \sigma_{j,m} = \lambda_m - \gamma k_j^2
                 = 4 c \sin^2\left( \frac{\pi m}{N} \right)
                   - \gamma \left( \frac{2 \pi j}{L} \right)^2
\end{equation}
Thus, the stability of the system depends on the sign of \(c\) as for \(j = 0\)
(the pure step bunching case) we have:
\begin{equation}
    \sigma_{0,m} = 4 c \sin^2\left( \frac{\pi m}{N} \right)
\end{equation}
If \(c > 0\), then there exist modes with \(m \ne 0\) such that
\(\sigma_{0,m} > 0\) and the system is unstable. Conversely, if \(c < 0\), then
all modes except the uniform translation mode decay since \(\sigma_{j,m} < 0\)
and the system is always stable to small perturbations. The mode \(j = m = 0\)
always has \(\sigma = 0\). The stability condition for the equally spaced step
train then is:
\begin{equation}
    \kappa_a p l_0^{-\left( p+1 \right)} - \kappa_r q l_0^{-\left( q+1 \right)} < 0
\end{equation}
or equivalently:
\begin{equation}
    \frac{\kappa_a}{\kappa_r} < \frac{q}{p} l_0^{p-q}
\end{equation}
This condition can be written more transparently as \(l_0 < w_M\), where:
\begin{equation}
    w_M \coloneqq \left( \frac{\kappa_r q}{\kappa_a p} \right)^{1/(q-p)}
\end{equation}
is the unique minimum of the step-interaction force
\(h(w) = \kappa_r w^{-q} - \kappa_a w^{-p}\). Thus instability occurs precisely
when the initial step spacing exceeds the ``natural'' spacing of the interaction
- the equally spaced train is linearly unstable for \(l_0 > w_M\) and stable,
apart from uniform translation, for \(l_0 < w_M\).

Further, one can deduce that the ``most dangerous'' mode is the
\(\sigma_{0, N/2}\)-mode, which is the antiphase bunching mode - steps start
moving exponentially fast away from one another, forming pairs with the step
behind them. This is seen in the numerical simulations as well - during the
surface evolution there is a tendency for steps to form and move in pairs
between bunches. Numerical solutions based on the Fourier series in
\autoref{eq:fourier_expansions} confirm that step-pairing is the dominant mode
in the linear regime. There is a slight subtlety here still - antiphase bunching
is the most dominant mode for even step counts \(N\), while for odd step counts
this maximum is achieved at the two neighboring modes
\(m = \left( N \pm 1 \right)/2\).

The importance of this result is twofold. First, the stability condition for
\(j = 0\) indeed recovers the stability condition obtained from analyzing the 1D
ordinary differential equations (ODE) based \textit{MM2} model of
\citep{Krasteva2016}. The \(\gamma\)-term in the model acts as a stabilizing
factor for high-wavenumber modes along the step edge, thus preventing the
development of arbitrarily small wavelength perturbations in the \(x\)
direction. The second reason this analysis is important is a more conceptual
one. The model presented in this way provides a ``smooth'' transition from 1D to
2D. More importantly, \autoref{eq:general_pde_form}, while constructed
\textit{ad hoc} for 2D, provides a general template to investigate all prior ODE
models as velocity functions and study their behavior in 2D. The \(\gamma\)-term
has an overall stabilizing effect on the step profiles, stabilizing them towards
straight steps, even in more complex cases where \(\gamma\) would depend on
temperature \citep{LWJeong1997}, miscut angle \citep{Stasevich2007}, lattice
structure, etc.

We note here that the linear stability analysis is under the assumption of small
perturbations. As the perturbations grow, the linearized model results diverge
from the true nonlinear dynamics of \autoref{eq:mm2_meandering}. For example, we
cannot capture long-time scale dynamics such as meander lifetime or step
coarsening using the linearized model. As perturbations grow the system goes
through a phase of intensive coarsening where the step edges are both meandered
and bunched. In \autoref{sect:structure} we already shed some light on the
nature of the long-term metastable meandered structures. They are purely (2+1)D
nonlinear phenomena which are not present in the model definition \textit{a
priori}, but rather are the result of the weak step stiffness not being able to
coarsen ``fast enough'' an already equilibrated in the step-normal direction
system. The \textit{MM2} gradient-flow nature admits a deeper variational
analysis of the equilibrium states of the original ODE model, motion of fronts
and the long-time coarsening of the meandered structures. These results are not
central to the morphological comparison aim of the current work and will be
reported separately with the necessary rigor of such analysis in
\citep{Ivanov2026}.

\section{Numerical solutions}

To further investigate \autoref{eq:mm2_meandering} we need to prepare an
appropriate numerical scheme to simulate the full nonlinear dynamics of the
system at arbitrary values of the parameters.

\subsection{Numerical scheme}

To allow for numerical stability through a wide parameter range we implement an
implicit finite difference scheme, where we discretize time as
\(t_{\mathfrak{i}} = \mathfrak{i} \tau\),
\(\mathfrak{i} = 0, 1, 2, \ldots, \mathbf{I}\) with time step \(\tau > 0\) and
the coordinate along the step edge as \(x_{\mathfrak{j}} = \mathfrak{j} h\),
\(\mathfrak{j} = 0, 1, \ldots, \mathbf{J}\) with spatial step \(h > 0\).

Let
\(y^\mathfrak{i}_{n, \mathfrak{j}} \approx u_n \left( t_{\mathfrak{i}}, x_\mathfrak{j} \right)\)
be the approximate numerical solution at time \(t_{\mathfrak{i}}\), then the
scheme stencil is given by:
\begin{align}
    \frac{y^{\mathfrak{i}+1}_{n, \mathfrak{j}} - y^{\mathfrak{i}}_{n, \mathfrak{j}}}{\tau} & = \gamma \frac{y^{\mathfrak{i}+1}_{n, \mathfrak{j}+1} - 2 y^{\mathfrak{i}+1}_{n, \mathfrak{j}} + y^{\mathfrak{i}+1}_{n, \mathfrak{j}-1}}{h^2} \notag                 \\
                                                                                           & - \kappa_a \left( \left( \Delta y^{\mathfrak{i}+1}_{n, \mathfrak{j}} \right)^{-p} - \left( \Delta y^{\mathfrak{i}+1}_{n+1, \mathfrak{j}} \right)^{-p} \right) \notag \\
                                                                                           & + \kappa_r \left( \left( \Delta y^{\mathfrak{i}+1}_{n, \mathfrak{j}} \right)^{-q} - \left( \Delta y^{\mathfrak{i}+1}_{n+1, \mathfrak{j}} \right)^{-q} \right)
\end{align}
We rearrange this to obtain the following nonlinear system for the unknowns
\(F_{n, \mathfrak{j}}^{\mathfrak{i}+1}(y) = 0\), such that:
\begin{align}
    F_{n, \mathfrak{j}}^{\mathfrak{i}+1}(y) & = y^{\mathfrak{i}+1}_{n, \mathfrak{j}} - y^{\mathfrak{i}}_{n, \mathfrak{j}} \notag - \gamma \frac{\tau}{h^2} \left( y^{\mathfrak{i}+1}_{n, \mathfrak{j}+1} - 2 y^{\mathfrak{i}+1}_{n, \mathfrak{j}} + y^{\mathfrak{i}+1}_{n, \mathfrak{j}-1} \right) \notag \\
                                            & + \tau \kappa_a \left( \left( \Delta y^{\mathfrak{i}+1}_{n, \mathfrak{j}} \right)^{-p} - \left( \Delta y^{\mathfrak{i}+1}_{n+1, \mathfrak{j}} \right)^{-p} \right) \notag                                                                                   \\
                                            & - \tau \kappa_r \left( \left( \Delta y^{\mathfrak{i}+1}_{n, \mathfrak{j}} \right)^{-q} - \left( \Delta y^{\mathfrak{i}+1}_{n+1, \mathfrak{j}} \right)^{-q} \right) \notag
\end{align}
Subject to periodic boundary conditions in \(\mathfrak{j}\) and the periodic
condition with the surface consistency constraint in \(n\):
\(y^\mathfrak{i}_{n+N, \mathfrak{j}} = y^\mathfrak{i}_{n, \mathfrak{j}} + Nl_0\)

We flatten the \(F_{n, \mathfrak{j}}^{\mathfrak{i}+1}(y)\) system into a vector
form to solve using a suitable Newton method. Since the Jacobian of the system
would be of size \(N \mathbf{J} \times N \mathbf{J}\), a matrix-free method
(GMRES by \citep{GMRES}) is used to solve the linear system at each Newton step
(i.e., a Newton-Krylov type nonlinear solver).

For the present work we implement the above scheme in Python using the JAX
library for automatic differentiation and GPU acceleration. Further, we use the
JAX-compatible Optimistix and Lineax libraries \citep{jax2018github, lineax2023,
optimistix2024} for the implementation of the Newton-Krylov solver. All arrays
are sharded along the step dimension \(n\) for efficient utilization of
multi-GPU systems and minimal device-to-device communication. This approach
makes the numerical method not only space-efficient (we do not need to store the
system Jacobian), but computationally efficient - we do not need to explicitly
calculate all the Jacobian entries, as JAX's autodiff capabilities let us
efficiently calculate Jacobian-vector products (JVP) only as required.

For the present work, systems of 50 steps with 1000 spatial nodes per step are
used. The matrix-free approach allows us to investigate these smaller systems
for long integration times without large computational resources. On the other
hand, the same scheme scales to larger systems of \(10^5\) steps easily when
needed.

All of the source code for the numerical scheme, including the configuration
files needed to reproduce the results in the current work, is provided as
supplementary materials to the text \citep{VassilCode}.

\subsection{Reconstruction of the 3D vicinal surface}

Although the model is (2+1)D and describes the motion of the step edges, those
edges can be viewed as the level-lines of the three-dimensional (3D) height
function \(h(t, x, y)\) with a height difference between two neighboring steps
of a fixed constant \(h_0\) - monostep height. This assumption is sensible as
the step-step repulsion in this model prevents the formation of macrosteps,
i.e., two or more steps stacked on top of each other.

Thus we can reconstruct the height function \(h(t, x, y)\) starting with
\(h(t, x, y) = 0\) at the first step and increasing the height by $h_0$ each
time a step is crossed. This follows Frank's \textit{ansatz} from his kinematic
theory \citep{Frank1959} which can be viewed as the continuum limit of the
above-described geometric procedure:
\begin{equation}
    \label{eq:frank_ansatz}
    \frac{\partial h}{\partial t} = - \frac{h_0}{\Delta u_n} \frac{\partial u_n}{\partial t}
\end{equation}

Given the above, we shall directly assume that we have a numerical approximation
of \(h(t, x, y)\) available for all subsequent analyses and comparisons.

\subsection{Numerical results}

All numerical results are obtained with the exponents set to \(p = 1\) and
\(q = 3\). The value of the repulsion exponent \(q\) corresponds to the elastic
potential \(n = 2\), i.e. \(q = n + 1\) from the original work of
\citet{Krasteva2016}, while \(p = 1\) is the simplest choice that produces an
attractive force. In the present work we will focus on this specific choice and
we will leave the investigation of the influence of these exponents on
morphologies for future works.
\begin{figure}[hbpt]
    \centering
    \includegraphics[width=0.85\linewidth]{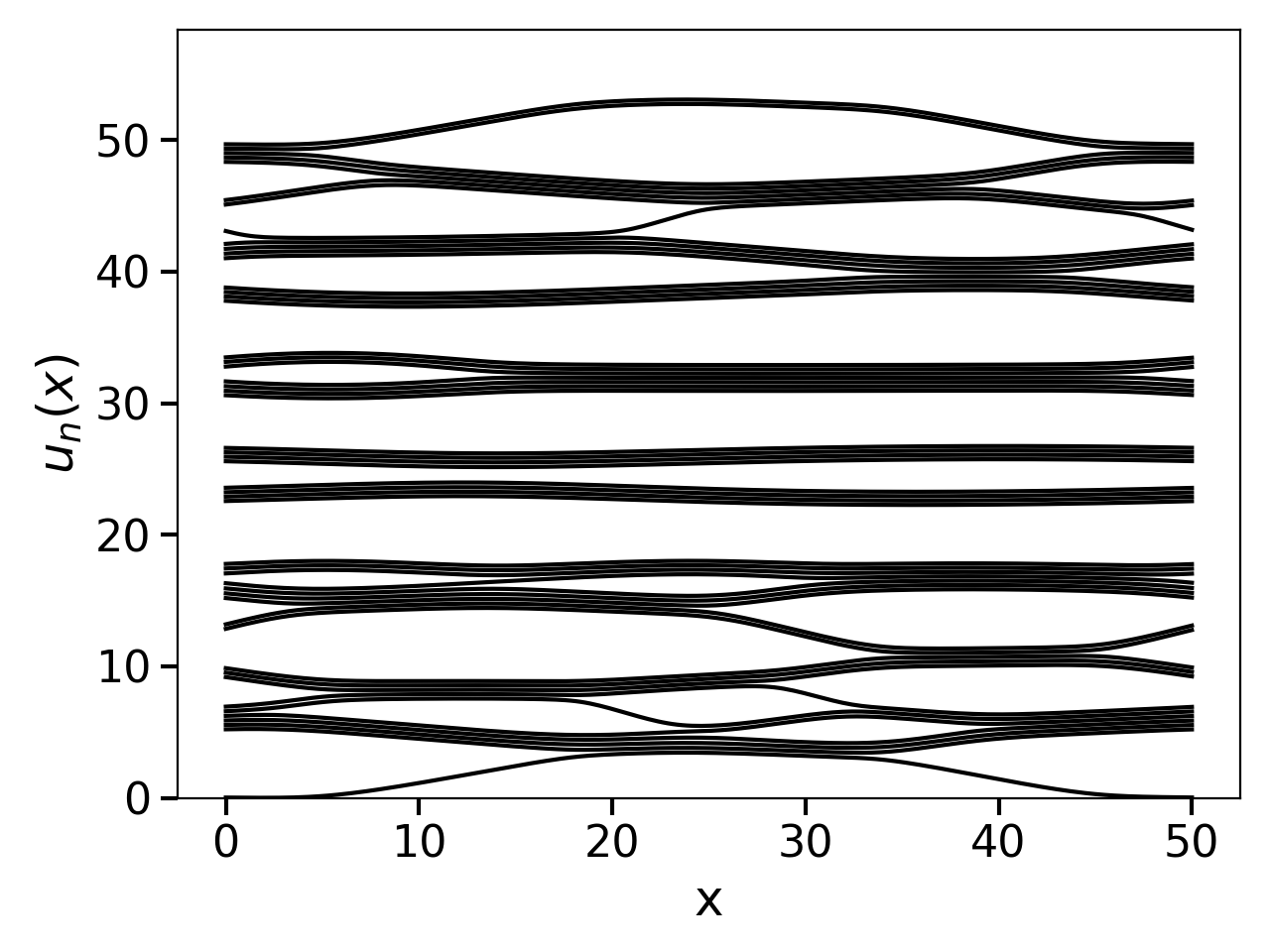}
    \caption{
        Early time snapshot of an already destabilized vicinal surface
        (\(N = 50\), \(L = 50\), \(\kappa_a/\kappa_r = 10\), \(\gamma = 50\),
        \(l_0 = 1\)) in the natural (2+1)D spatial setup. The wavy curves
        represent the step positions. Note that steps do not cross, but only
        meander and/or bunch and there is a tendency for paired steps to move
        together. Further, the minimal step to step distance within the bunch
        does not depend on the number of steps in the given bunch.
    }
    \label{fig:2d_fig_continuum}
\end{figure}

\begin{figure}[hbpt]
    \centering
    \includegraphics[width=0.45\linewidth]{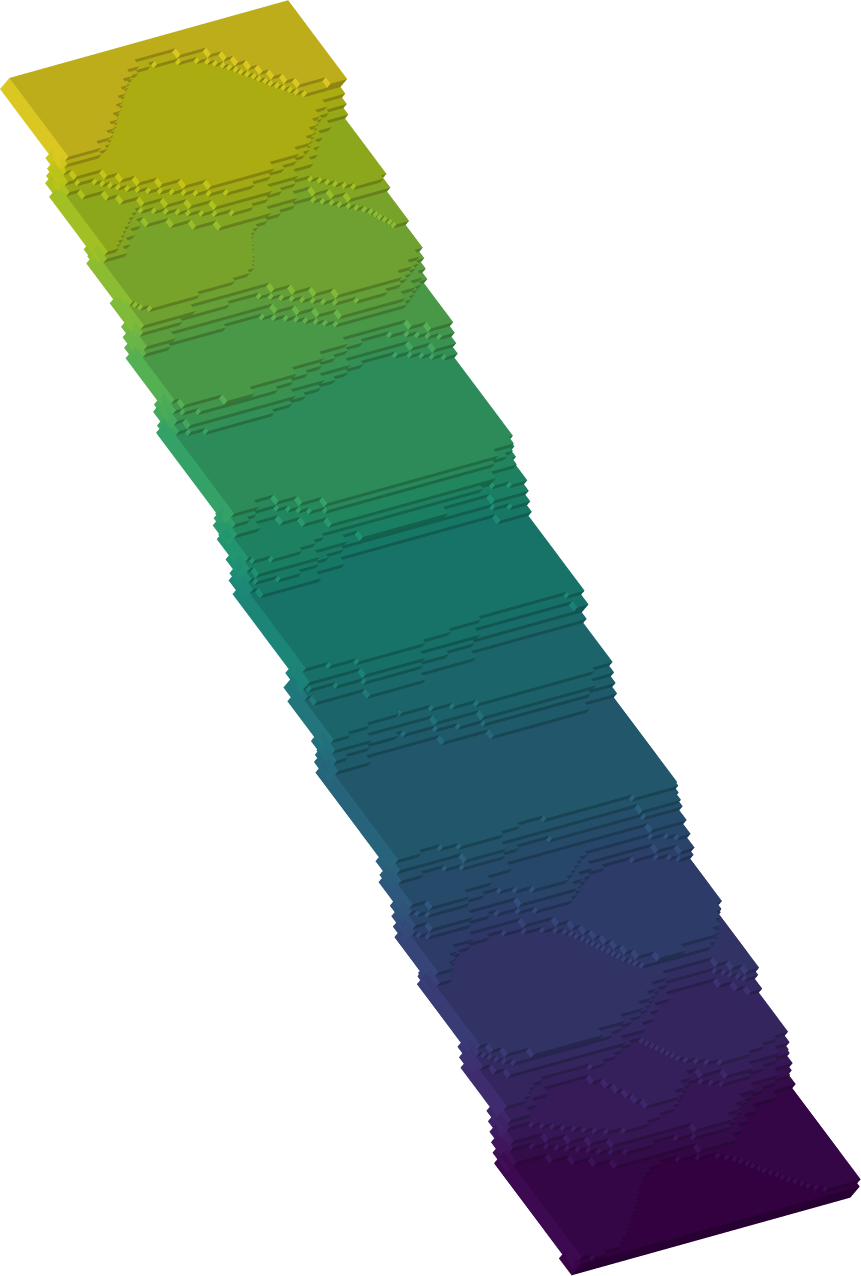}
    \caption{
        Reconstruction of the 3D vicinal surface based on
        \autoref{eq:frank_ansatz} of \autoref{fig:2d_fig_continuum} (\(N = 50\),
        \(\kappa_a/\kappa_r = 10\), \(\gamma = 50\)). Note the simultaneous
        occurrence of both meandering and bunching. Each terrace is color-coded
        according to its height \citep{ovito}.
    }
    \label{fig:3d_reconstruction}
\end{figure}

As K\&W have noted in \citep{Kandel1994} this simple model is able to produce a
surprising range of morphologies in which bunching, meandering, bunching along
with meandering, are all possible. \autoref{fig:2d_fig_continuum} and
\autoref{fig:3d_reconstruction} present the early stages of one such case. This
single case already demonstrates the richness of behaviour the model can
present.

\begin{figure*}[hbt]
    \centering
    \includegraphics[width=0.95\linewidth]{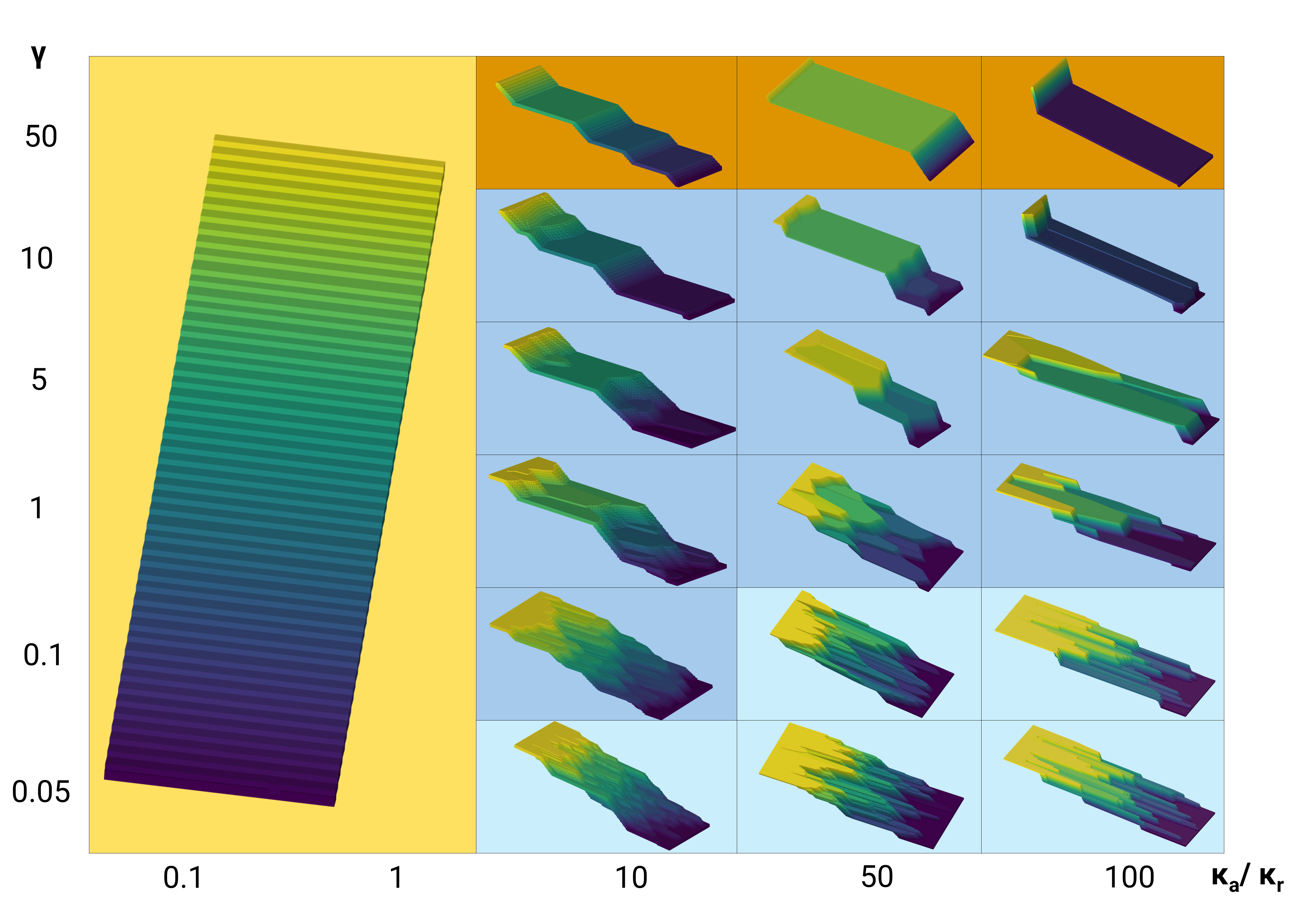}
    \caption{
        Structures obtained for \(N = 50\), \(L = 50\), \(l_0 = 1\), \(p = 1\),
        \(q = 3\) as a function of the control parameters \(\kappa_a/\kappa_r\)
        and \(\gamma\) for long simulation time \(T = 50\) from
        \autoref{eq:mm2_meandering}. Steps are discretized with 1000
        equally-spaced nodes and the time step is \(\tau = 10^{-4}\).Different
        background colors denote regions with different patterns: yellow -
        straight steps, orange - straight step bunching, blue - simultaneous
        bunching and meandering and light blue - step meandering.
    }
    \label{fig:3d_diagram}
\end{figure*}

We further construct a morphological diagram in \autoref{fig:3d_diagram} that
shows an exploration of the 2D-parameter space comprised of \(\gamma\) on one
axis and \(\kappa_a/\kappa_r\) on the other that directly illustrates how the
system morphologies change as the parameters are changed. Furthermore, we
provide 3D video clips of the full system dynamics as supplementary material to
the present text.

From \autoref{fig:3d_diagram} clear regions (shaded in different colors) in the
diagram can be noticed - straight step bunching (orange background) at high
\(\gamma\), bunching + meandering (blue background) at intermediate values of
\(\gamma\) and ``pure'' meandering (light blue background) at low \(\gamma\).
The destabilizing force of \(\kappa_a/\kappa_r\) makes each of the structures
more pronounced - bunches become more tightly packed (i.e. \(l_{min}\) - the
minimal step-step distance in the bunch decreases) and meanders become more
``extreme''. The B1-type of the \textit{MM2} model is preserved here, in the
pure step bunching case and in the case of bunched meanders, the minimal
distance in the bunch \(l_{min}\) does not depend on the number of steps \(N\) -
the bunches are incompressible, which is typical and expected of heteroepitaxial
B1-type models. The preservation of this model property is not \textit{a priori}
clear from \autoref{eq:mm2_meandering} and poses the question of the
preservation of the scaling of \(l_{min}\) with the number of steps in the bunch
for other choices of velocity functions.

Finally, the low-pass filter effect of \(\gamma\) can be observed in the
wavelength of the meanders. At higher values, only the lower-frequency
meandering modes survive and define the structure.

\section{Morphologies generated by the Vicinal Cellular Automaton model}

\subsection{The VicCA model}

\begin{figure}[hbt]
    \centering
    a)
    \includegraphics[width=0.44\linewidth]{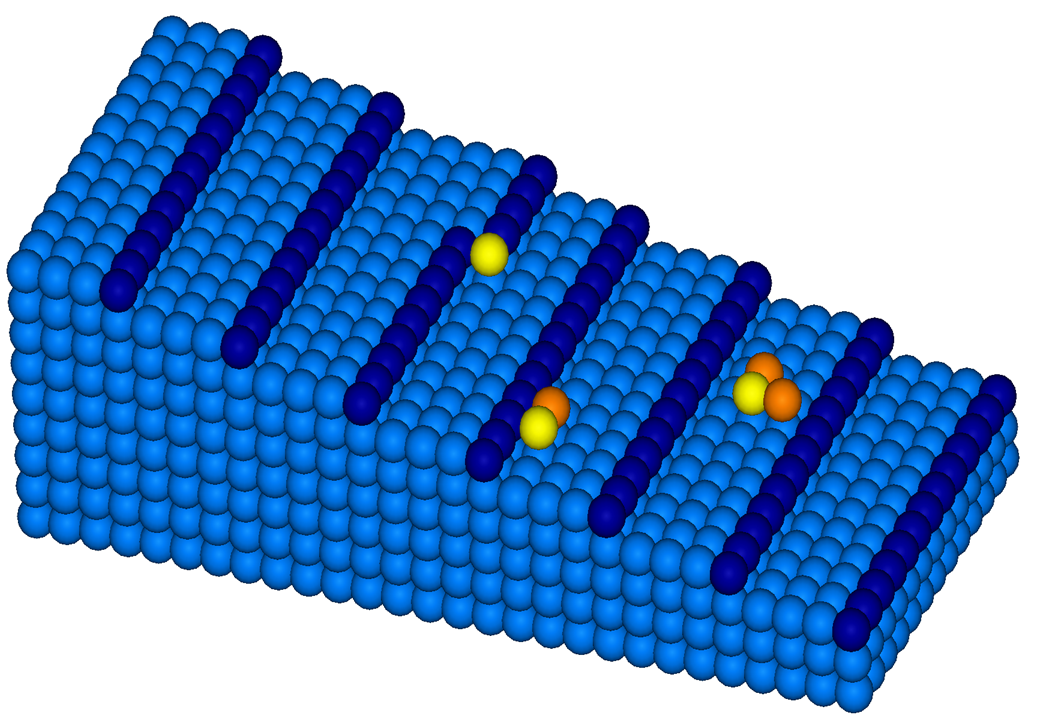}
    b)
    \includegraphics[width=0.44\linewidth]{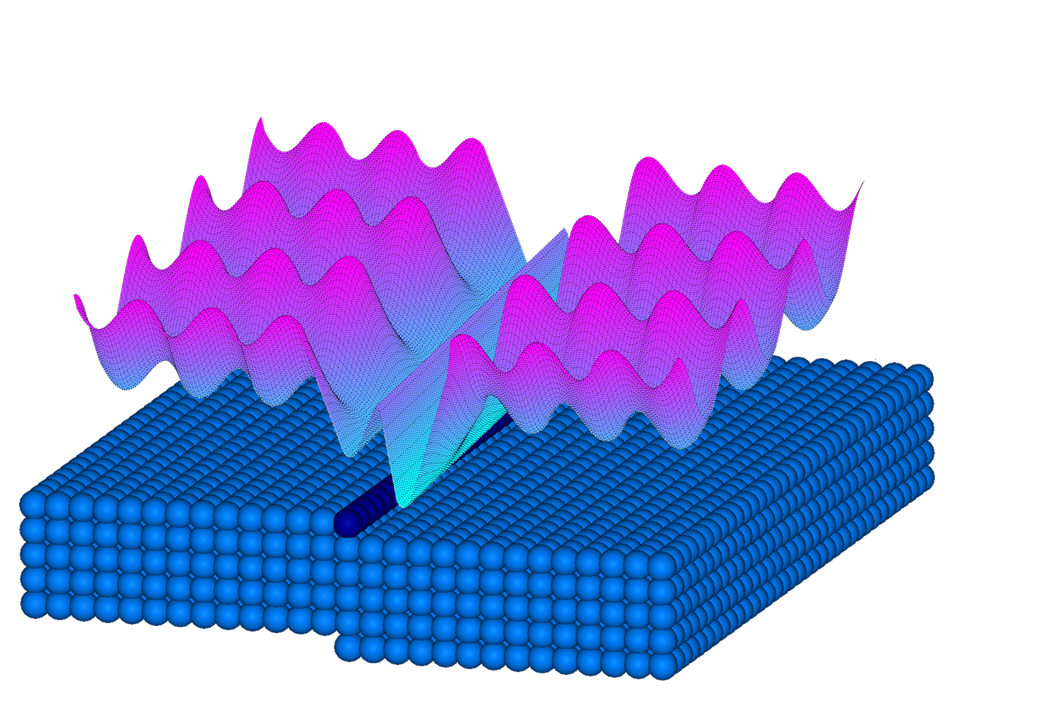}
    \caption{
        a) Crystal structure with the diffusing adatoms on the surface. b) Shape
        of potential energy in which diffuse adatoms.
    }
    \label{fig:model}
\end{figure}

The surface evolution is now modeled using a (2+1)D Vicinal Cellular Automaton
(VicCA) framework that combines automaton rules, accounting for the ordering
role of crystal growth, with a Monte Carlo module describing the disordered
diffusion field through the fine details of the adatom distribution
\citep{zaluska-kotur_step_2021, Chabowska-JCG, Chabowska-PRB,
chabowska_surface_2025}. The model represents a Kossel crystal lattice and a
surface layer of mobile adatoms on top of it that incorporate into the crystal
during the growth updates (see \autoref{fig:model}a). The growth \textit{rules}
are defined only once - at the beginning and could be easily changed from
experiment to experiment to trace the impact of the model parameters. The
parallel fashion in which the automaton is updated according to the growth rules
is superior to the classical kinetic Monte Carlo simulations, like those
previously employed to study the bunching and meandering processes of SiC
crystals \cite{Krzyewski2014} where each local change of a system state is valid
immediately and impacts the neighboring state further. The VicCA is also
straightforward in that no dimensional quantities are used - thus the system
trajectories are universal (for a given set of dimensionless parameters).
Performing all possible growth events at once permits us to study different
regimes of the diffusion rate by varying the number of diffusional updates,
\(n_{DS}\), which determines how many diffusional hops each adatom attempts on
average between growth updates.

The crystal is defined as a vicinal crystal surface - a miscut from a Kossel
crystal, with a 3D cubic lattice, consisting (initially) of monoatomic steps
with periodic boundary conditions along the steps and helical boundary
conditions across them. The geometry and properties of the substrate layer do
not differ from the subsequently grown layers, thus serving as an archetypical
example of homoepitaxy (or, simply, crystal growth). The most favored by the
rules incorporation sites are the step voids and kinks - where the attachment to
the crystal is unconditional, whereas less coordinated sites require additional
conditions. For example, the attachment to a straight step, the so called 1D
nucleation, requires the presence of a neighboring adatom adjacent to the
attaching one, thus mimicking the attachment to the kink, but with an adatom in
the configuration. The ratio of adatom attachment probabilities at kink and step
positions effectively controls how straight the step is. In other words, the
step-edge growth is treated as a nucleation process with a critical nucleus size
of two atoms.

Adatom diffusion serves as the primary driving mechanism for surface pattern
formation. In the model, adatoms diffuse with transition probabilities governed
by local energy barriers. They migrate across terraces by hopping between
adsorption sites, overcoming the corresponding diffusion barriers. Adatom jumps
are temperature-activated, with the probability
\begin{equation}
    P_J = P_0\exp(-\beta\Delta V)
\end{equation}
where $\Delta V = V_f - V_i$ is the potential energy difference between the
final and initial sites, $\beta = (k_BT)^{-1}$ is a temperature factor and $P_0$
is an overall normalization factor that sets the time scale.

The energy of atoms at adsorption sites is uniform across the terrace, except in
the vicinity of steps. In the present study, this potential energy landscape is
extended by introducing additional energy wells at both the top and the bottom
of each step. The well at the top of the step constitutes a novel extension
compared to \citep{Chabowska-PRB}. Consequently, two potential wells are now
present at each step edge (see \autoref{fig:model}b), generating a
position-dependent energy landscape that evolves alongside the step morphology.
The depths of these wells, $E_V$ at the bottom and $E_S$ at the top of the step,
serve as key control parameters that determine the resulting surface ordering.
Because the shape of this energy landscape plays a crucial role in surface
dynamics, and various potential profiles may arise, we examine in the following
section how the relative depths of these wells influence surface pattern
formation. Both sites—at the top and bottom of the steps—are distinct, as
adatoms occupying them interact differently with the crystal atoms compared to
those on the rest of the terrace. Such adatoms may form additional bonds or
experience bond modifications due to surface step reconstruction. By adjusting
the potential energy wells at these locations, we explicitly account for these
effects. The adatom layer diffusing across the crystal surface moves from step
to step, adapting to a step-dependent local potential landscape. In this way,
information is transferred between steps. For example, when steps are close
together, they capture adatoms, leading to a lower local density and slower step
motion. In general, the precise dependence between step density and step motion
is closely tied to the potential shape and can be translated into the step-step
interaction terms. This relation is analyzed below.
At each simulation step, diffusion is followed by a growth update via CA rules,
after which the adatom concentration is replenished to a fixed value by adding
adatoms uniformly at random over the diffusion layer, mimicking a constant
external incoming flux of particles. Unlike conventional Monte Carlo models,
VicCA decouples diffusion, attachment, and nucleation processes, allowing
independent control of kinetic parameters such as step attachment, kink
incorporation, diffusion length, and external flux. In CA simulations, all
potential energies are measured in units of thermal energy $k_BT$. The time and
temperature scales are set through the diffusion barriers, the number of
diffusion attempts, and the adatom concentration, providing a flexible framework
for studying step bunching, meandering, and more complex patterns arising from
their interplay. Note that due to the discreteness of the adatom layer, unlike
in continuum models, attachment to the step occurs at randomly distributed
locations. Consequently, no initial perturbation is required to initiate the
process or destabilize the step positions.

\subsection{Comparing the models}

With the VicCA lattice-based atomistic model defined, we can now analyze it by
comparison with the continuum framework presented previously. We consider its
general differential form, given by \autoref{eq:mm2_meandering}. This equation
relates the velocity of the \(n\)-th step to the step stiffness, expressed by
the second derivative of the step profile with respect to the \(x\) direction,
and to two additional terms representing step–step interactions, namely the
attractive and repulsive contributions. These interaction terms share the same
functional form and are expressed as a power of the terrace width. Altogether,
the model depends on five parameters. The first is the stiffness parameter,
$\gamma$. Two additional parameters, $p$ and $q$, determine the power law
dependence of the step–step interactions on the terrace width. Finally, the
coefficients $\kappa_a$ and $\kappa_r$ describe the strengths of step–step
attraction and repulsion, respectively.

The bridge between $\gamma$ and the potential well energies shall be obtained
through the average deviation of the meandered step from its average position.
Specifically, we are interested in the mean square
width:\begin{equation}
    \langle w^2 \rangle = \langle u_n^2(t,x) \rangle_x
                          - \langle u_n(t,x) \rangle_x^2
\end{equation}
where the operator $\langle f(x)\rangle_x$ denotes the spatial average of the
quantity $f(x)$ over $x$. Following the work of \citep{Pimpinelli1993,
Bartelt1993, Ihle1998, Misbah2010}, it is established that the mean square
displacement $\langle w^2 \rangle$ over a distance $L_w$ scales as:
\begin{equation}
    \frac{\langle w^2 \rangle}{L_w} \propto \frac{k_B T}{\tilde \beta}
\end{equation}
where $\tilde \beta$ is the step stiffness (with dimensions of energy per unit
length) as defined in \autoref{eq:main_hamiltonian}, and $k_B T$ represents the
thermal energy. For a meandered step, we consider the distance of the wavelength
$L_w = \lambda$. We observe that at the initial stage of the bending process,
the fluctuation is on the order of the lattice unit distance, i.e.,
$\langle w^2 \rangle \approx a^2$. Furthermore, based on the chemical potential
reservoir construction in \citep{LWJeong1997}, we have:
\begin{equation}
    \gamma = \frac{\Gamma_A \tilde \beta}{k_B T}
\end{equation}
where $\Gamma_A$ is the step mobility constant. By combining these relations, we
establish that:
\begin{equation}
    \gamma \propto \frac{\Gamma_A \lambda}{a^2}
\end{equation}
In previous work, \citep{Chabowska-PRB} demonstrated that the meander wavelength
satisfies:
\begin{equation}
    \label{eq:vicca_meander_wavelength}
    \lambda \propto \frac{\rho_k}{\rho_{+}^2}
\end{equation}
while the forward velocity of the step is given by $\Gamma_A = \rho_+^2$. In
this formulation, $\rho_k$ represents the adatom density at the kink position,
while $\rho_+$ and $\rho_-$ denote the densities immediately adjacent to the
step edge. Specifically, $\rho_+$ is the density in front of the step on the
lower terrace, expressed as:
\begin{equation}
    \rho_+ \propto c_0 e^{\beta E_V}
\end{equation}
where $c_0$ is the average adatom concentration on the terraces and
$\beta = (k_B T)^{-1}$. Correspondingly, the density on the upper terrace is
given by $\rho_- \propto c_0 \exp(\beta E_S)$. The local concentration at the
kink, $\rho_k$, is determined by the flux of particles arriving from neighboring
sites; once an adatom transitions to a kink site, it is incorporated into the
crystal lattice. We can therefore evaluate $\rho_k$ as a weighted sum of the
concentrations at these adjacent sites. Since this sum is primarily dominated by
contributions from the upper and lower terraces, the relationship can be
approximated as:
\begin{equation}
    \gamma
    \propto \rho_k
    \approx \frac{c_0}{2 a^2} \left( e^{\beta E_V} + e^{\beta E_S} \right)
\end{equation}
This expression effectively links the microscopic attachment energies to the
macroscopic line tension and step stiffness.
\begin{figure*}
    \centering
    \includegraphics[width=0.90\textwidth]{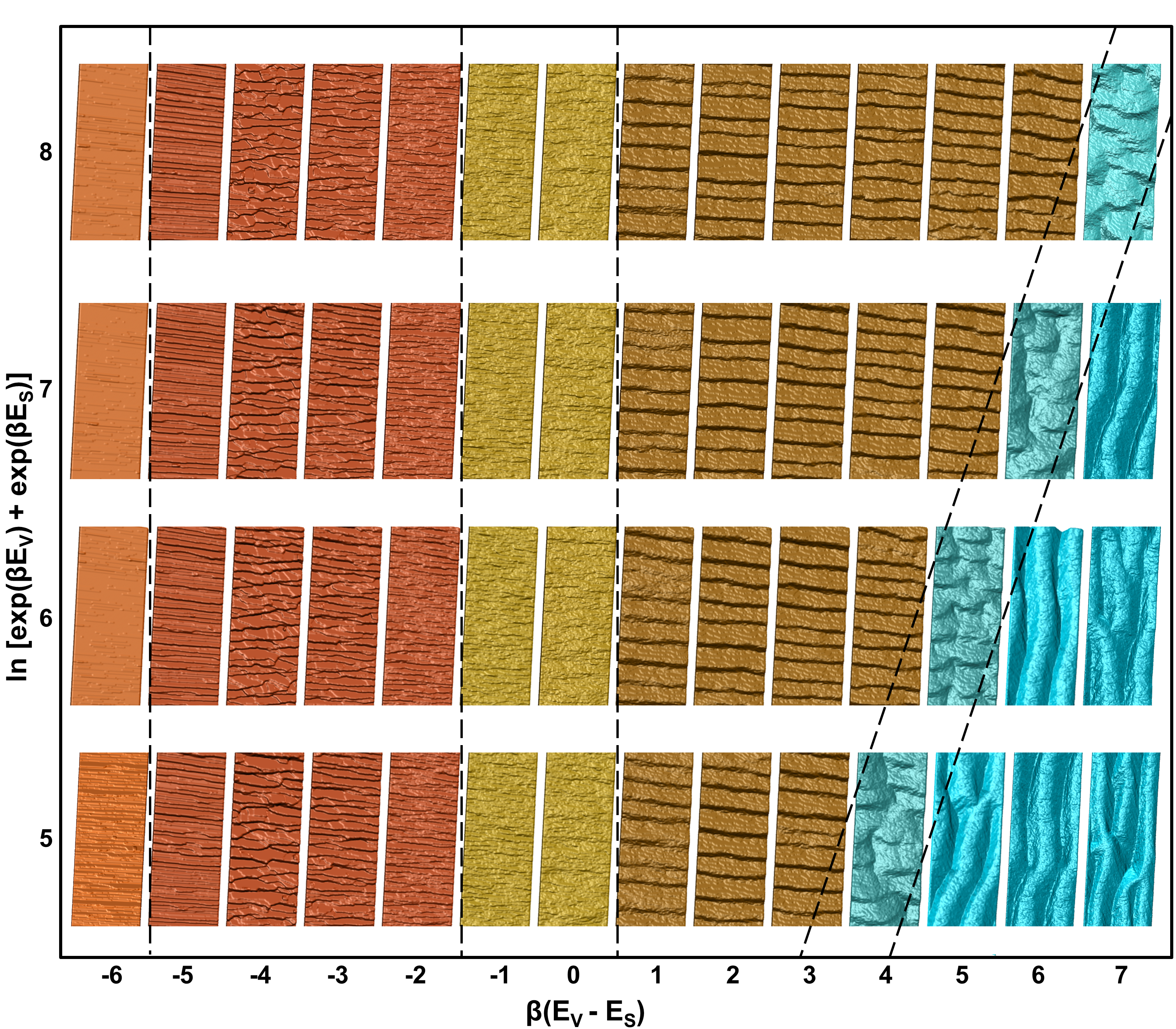}
    \caption{
        Structures obtained for $n_{DS} = 30$, $c_0 = 0.02$, $l_0 = 4$, as a
        function of the depths of the potential wells $E_S$ on top and $E_V$ at
        the bottom of the step. Simulation time $10^6$. System size
        $1000 \times 200$.
    }
    \label{fig:simulations1}
\end{figure*}

More subtle, but crucial for the dynamics, is the dependence of this density on
the widths of the terraces on both sides of the step. This terrace width
dependence provides the dominant contribution to the effective step–step
interaction terms, which in turn govern the onset of step instabilities. From
the atomistic simulation perspective, two main mechanisms lead to step
instabilities. The first is the net flux of adatoms descending from the upper
terrace during crystal growth. As the width of the upper terrace increases, a
larger number of adatoms reaches the step edge, enhancing the step velocity.
This mechanism promotes step bunching and represents a global instability driven
by mass transport asymmetry. As a result, it acts as an effective attractive
interaction between steps. The other mechanism is linked to the gathering of
adatoms close to the step, which pushes the step forward. The local increase in
adatom density depends on the depth of the potential well and on the terrace
width. The wider the terrace below the step, the larger the density in the
potential energy well, the faster the step moves; effectively, when we subtract
the mean step motion corresponding to the average terrace width, this
contribution can be described by a step–step repulsion term.

\begin{figure}[h!]
    \centering
    \setlength{\tabcolsep}{3pt}
    \renewcommand{\arraystretch}{0.86}
    \begin{tabular}{|l c c|}
        \hline
        a) &    \includegraphics[width=0.41\linewidth]{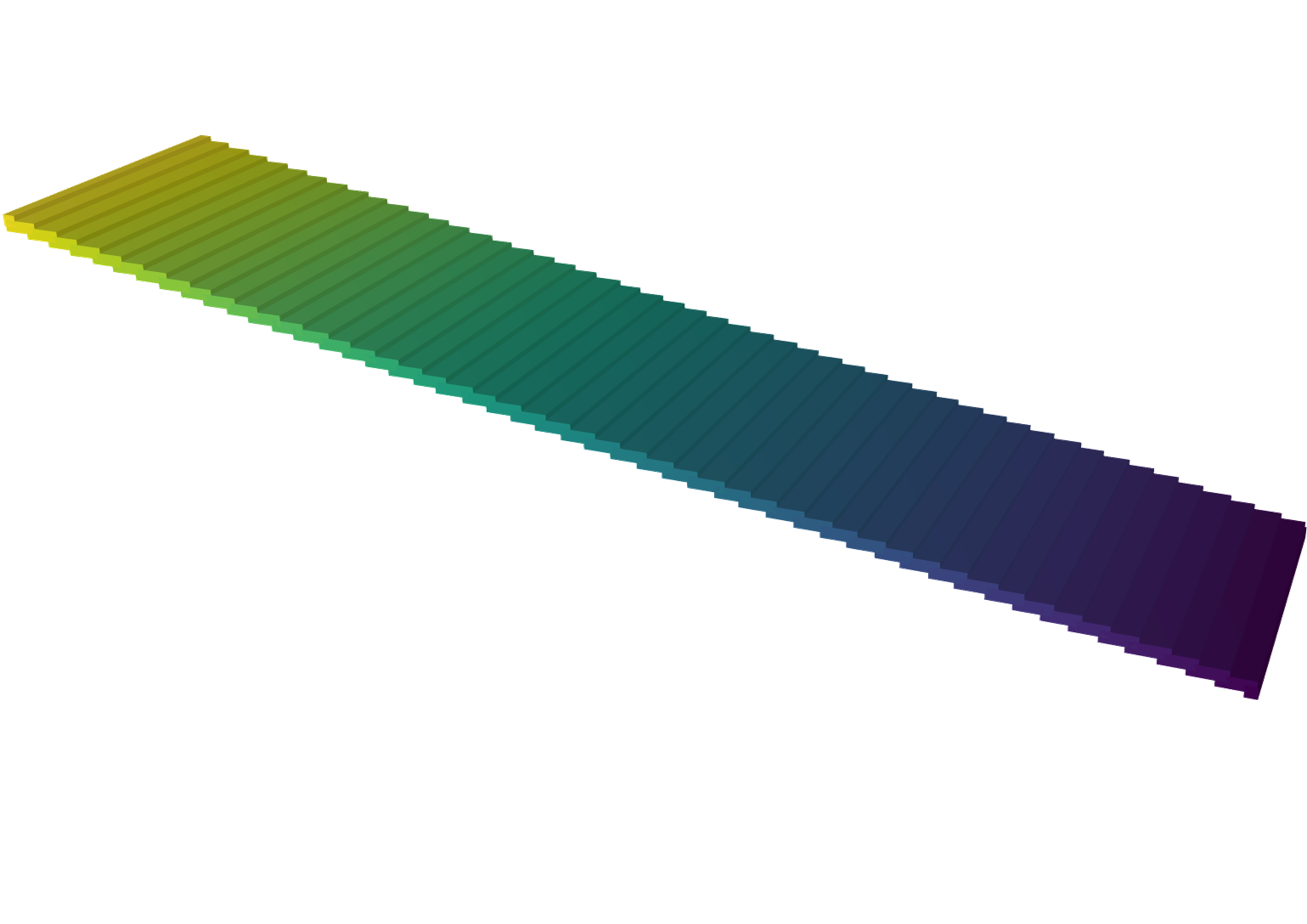}    & \includegraphics[width=0.49\linewidth]{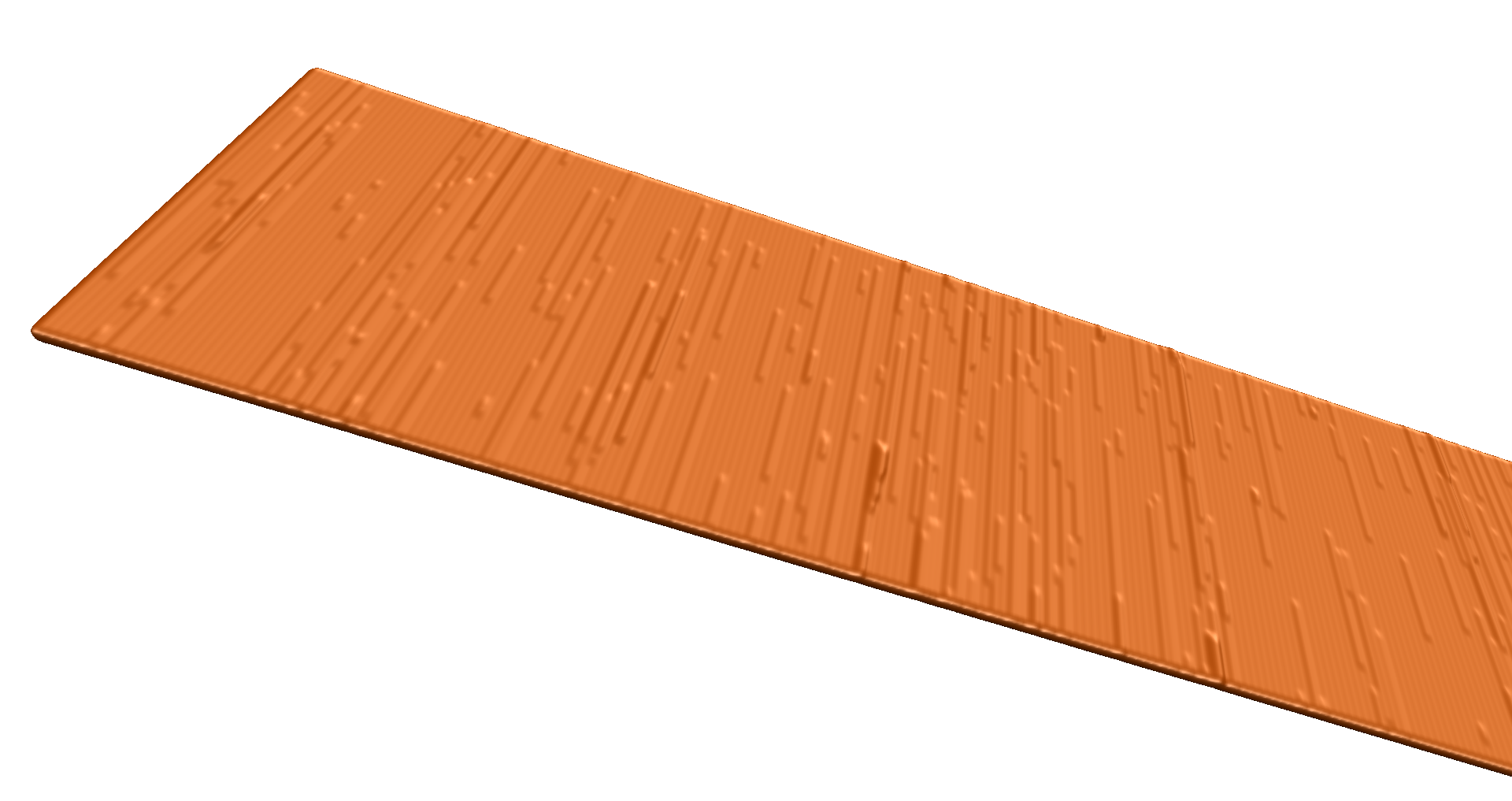} \\
        \hline
        b) &     \includegraphics[width=0.42\linewidth]{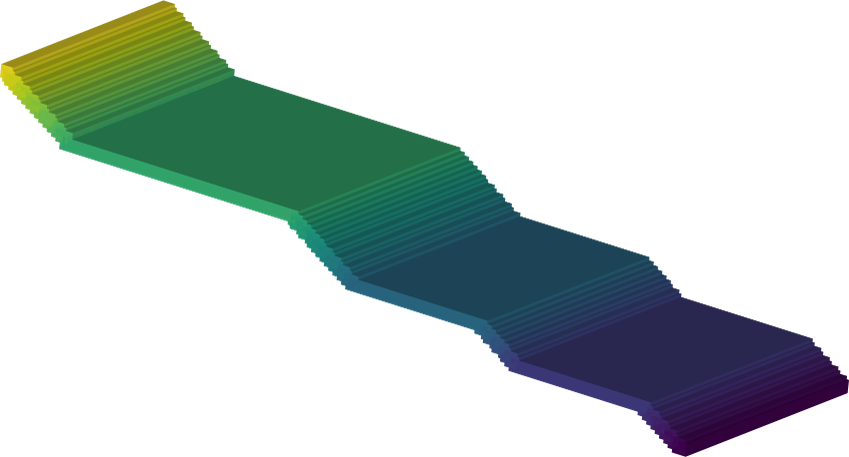}     & \includegraphics[width=0.47\linewidth]{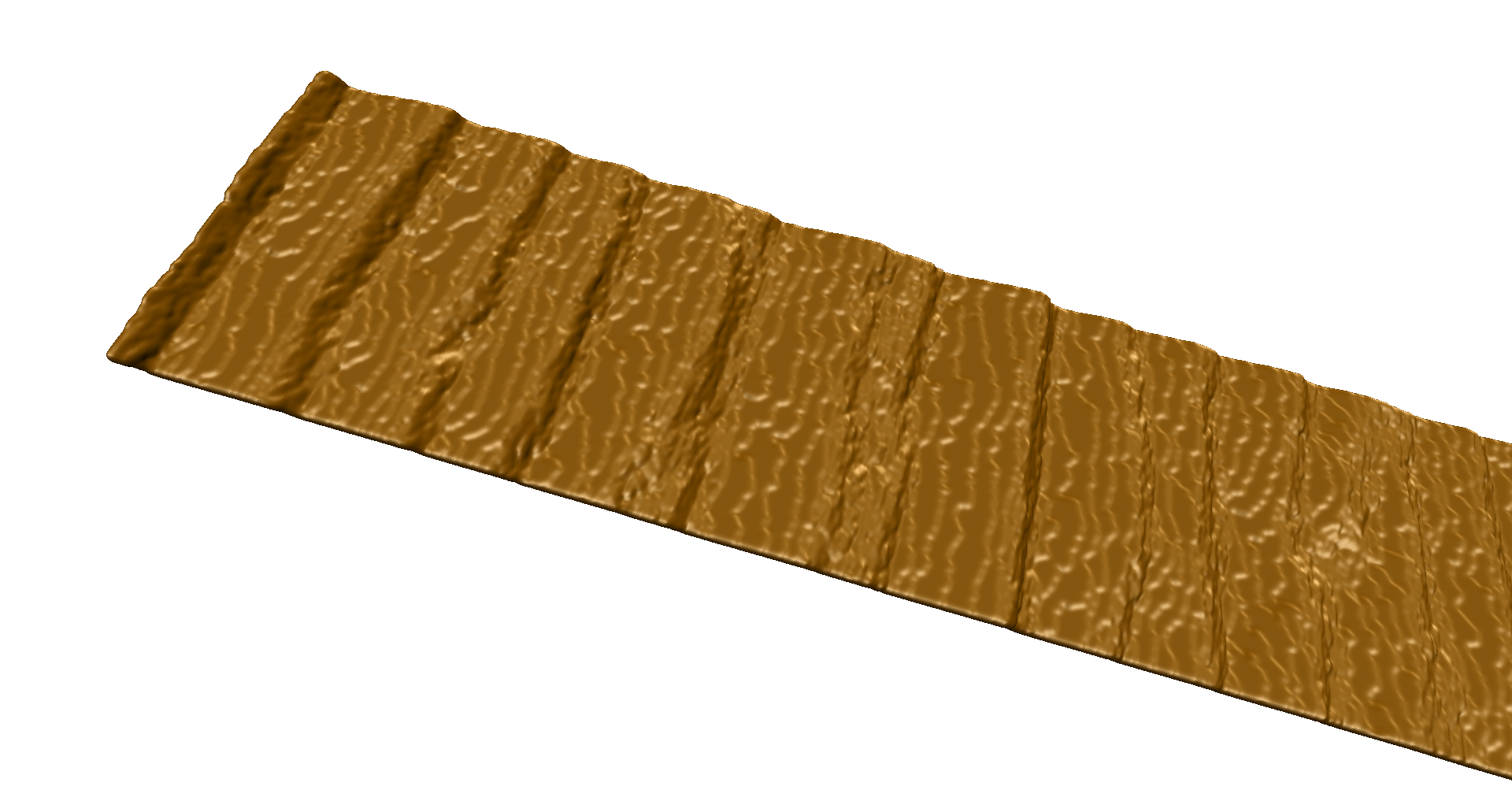} \\
        \hline
        c) & \includegraphics[width=0.42\linewidth]{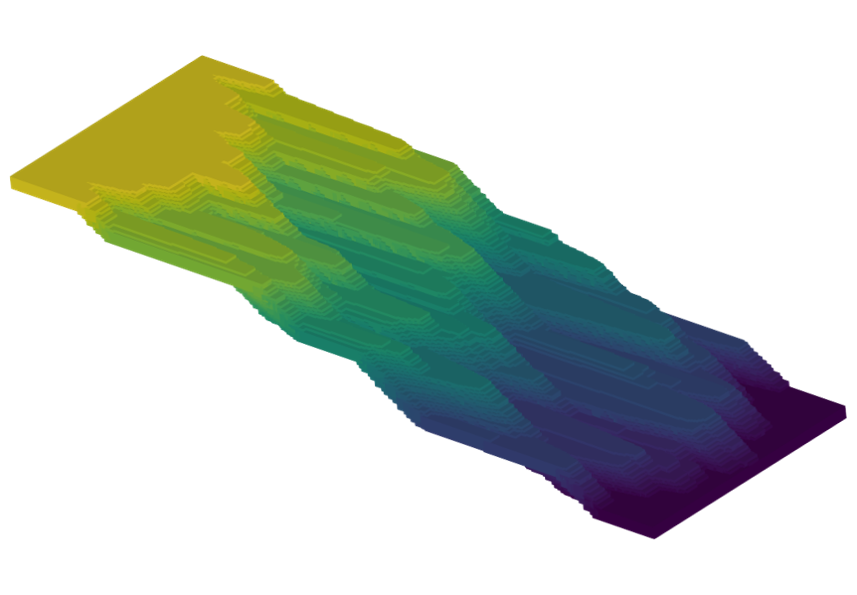} & \includegraphics[width=0.49\linewidth]{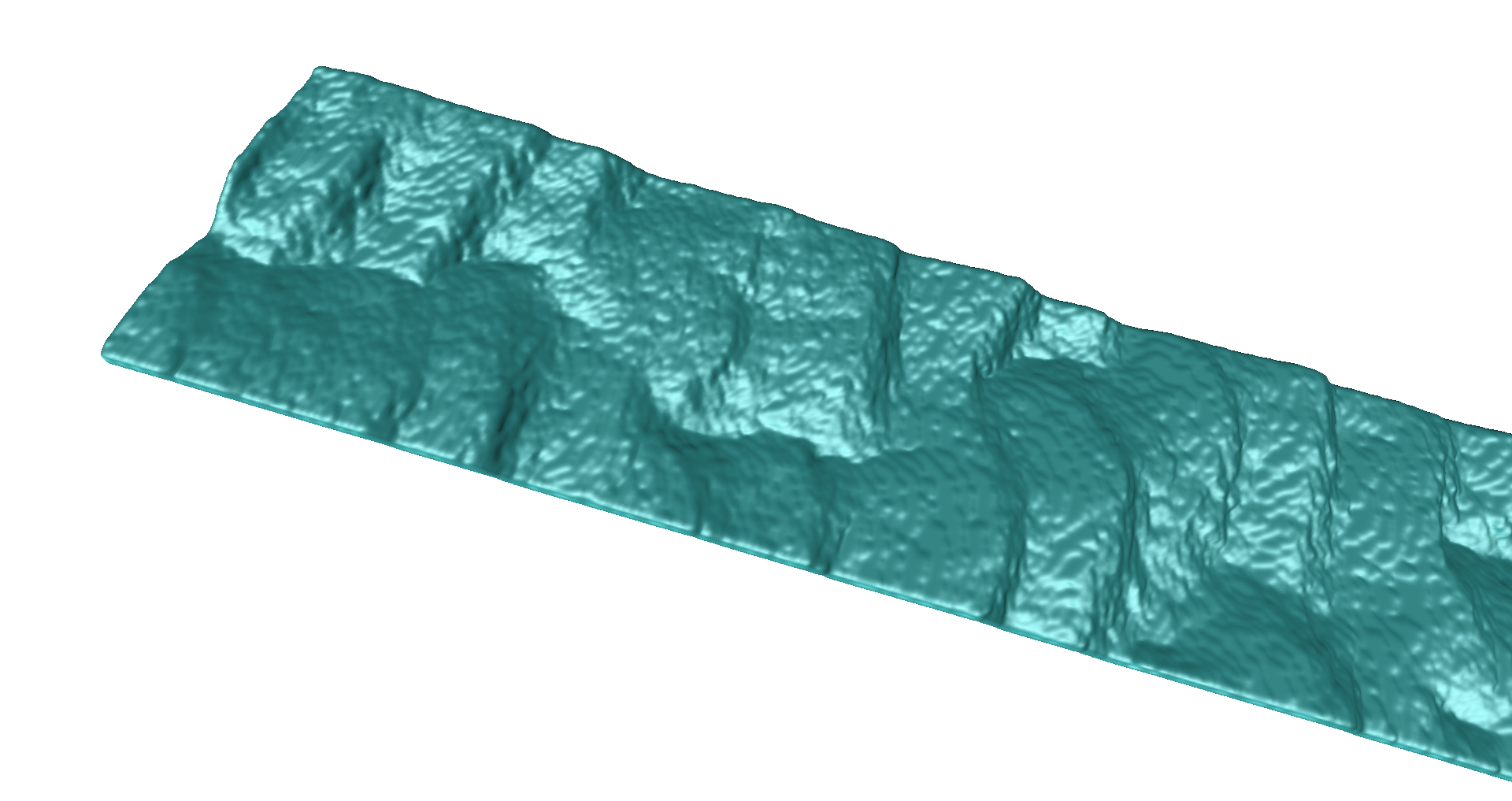} \\
        \hline
        d) &  \includegraphics[width=0.42\linewidth]{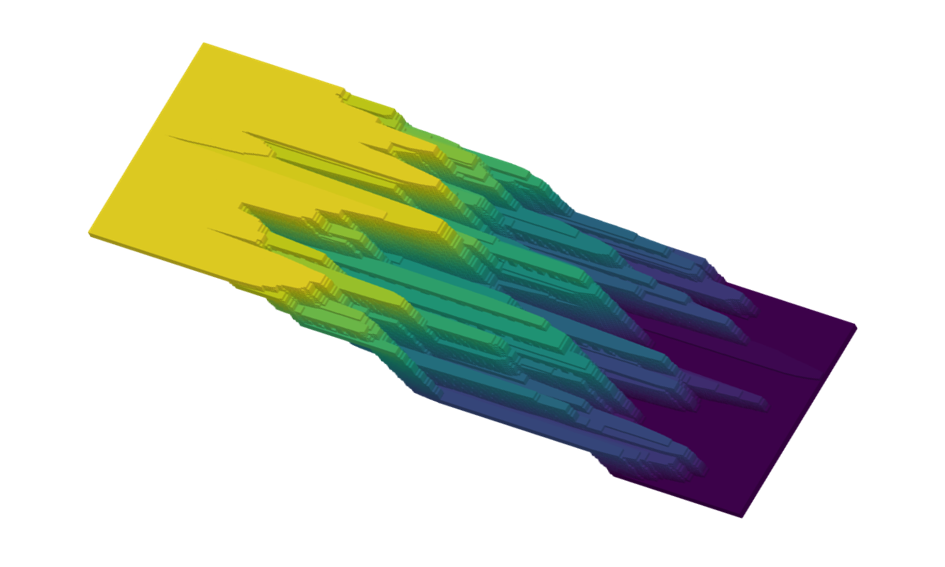}  & \includegraphics[width=0.49\linewidth]{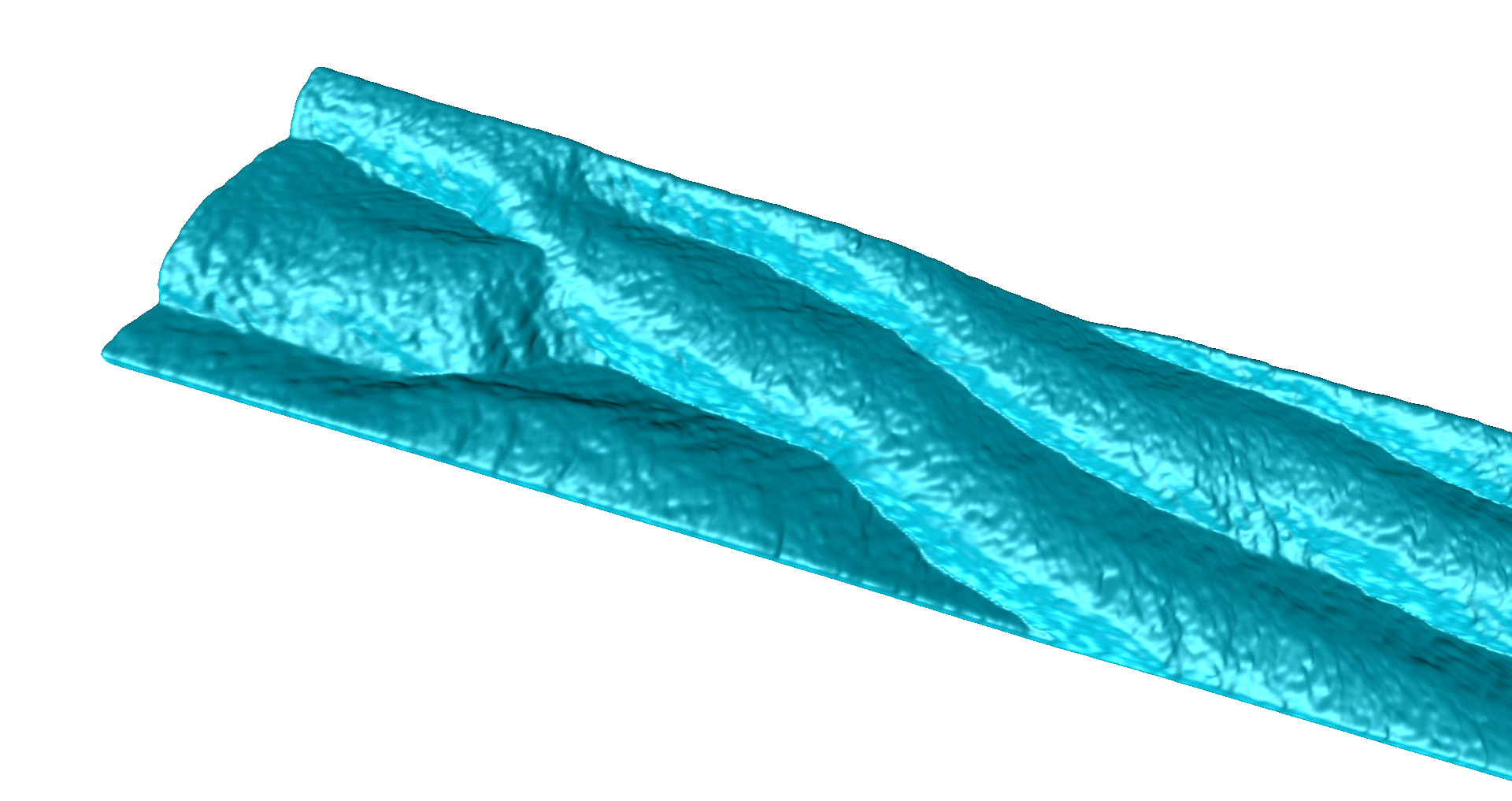} \\
        \hline
    \end{tabular}
    \caption{
        Comparison of surface morphologies obtained from the pattern diagram in
        \autoref{fig:3d_diagram} (left) and numerical simulations in
        \autoref{fig:simulations1} (right). The panels illustrate: (a) regular
        step trains, (b) step bunching, (c) simultaneous bunching and
        meandering, and (d) step meandering.
    }
    \label{fig:comparison}
\end{figure}

The qualitative tendencies are clear, although the appropriate powers $p$ and
$q$ to be used here are less obvious. Nevertheless, it is certain that the
presence of a potential well at the step bottom leads to step meandering
\citep{Chabowska-PRB}. Specifically, as the well depth increases, the
wavelengths of the meanders decrease. This can be explained by a reduction in
step stiffness \autoref{eq:vicca_meander_wavelength}. In fact, any modification
to the local potential energy influences the adatom density at the top or bottom
of the steps, which subsequently affects step kinetics. The resulting step
motion can be effectively translated into step stiffness or step-step
interactions, both of which are dependent on local densities. In the limit of
high potential, this density change at the bottom of the step can be
approximately expressed as $c_0\exp(\beta E_V)$, which in turn is proportional
to the increase in the step velocity. For linear perturbation of
\autoref{eq:mm2_meandering}, it can be equated to $\kappa_a p$, expressing the
attractive step-step interaction. Meanwhile, the change of the density due to
the potential at the top of the step - proportional to $c_0\exp(\beta E_S)$ -
gives the repulsive forces and can be equated to $\kappa_r q$. Thus,
\({\kappa_a p / \kappa_r q \approx \exp( \beta (E_V - E_S))}\).

These relationships should be interpreted as an effective parameter
correspondence and not as an exact ``homogenization'' between the micro- and
meso-scale models. The continuum model coarse-grains the surface diffusion and
the attachment mechanisms into the three effective quantities
\(\gamma, \kappa_a, \kappa_r\), while the VicCA retains the discrete nature of
attachment events. Consequently the models are compared on the level of
morphology classes, instability types and high-level correspondence of control
parameters, rather than specific point-wise correspondence of the results.

We now analyze the evolution of step patterns as a function of two distinct
combinations of the potential wells, $E_V$ and $E_S$ (\autoref{fig:model}b). The
resulting surface patterns are presented in \autoref{fig:simulations1}, mapped
against $\beta (E_V - E_S)$ on the horizontal axis and
$\ln[\exp(\beta E_V) + \exp(\beta E_S)]$ on the vertical axis. The phase diagram
is divided into distinct regions where similar surface morphologies emerge, with
dashed lines indicating the boundaries between these regions. Considering the
right part of the diagram, in the bottom right corner, regular and pronounced
elongated meanders, referred to as ``finger-like'' structures, are formed. This
morphology closely resembles the results obtained from the continuum model
(\autoref{fig:3d_diagram}). Moving upward, a transition occurs toward a region
of mixed morphology, featuring both step meandering and bunching. At even higher
values, the surface evolves into a fully bunched morphology, marking a clear
shift in the dominant instability. The left part of \autoref{fig:simulations1}
contains three regions, marked by different colours. The VicCA results show
regular step trains at the far left, small bunches of slightly bent steps, and
more irregular step edges closer to the center. In the corresponding region of
\autoref{fig:3d_diagram}, the continuum model produces only regular trains of
straight steps. \autoref{fig:comparison} compares examples from the two
morphology diagrams: (a) regular step trains, (b) step bunching, (c)
simultaneous bunching and meandering, and (d) step meandering. The VicCA example
in panel (a) is taken from the far left of \autoref{fig:simulations1}. Both
models reproduce these morphologies, although the bunches in the VicCA examples
are less pronounced than those in the continuum results. This reflects the
limited duration of the VicCA simulations and the different timescales of the
two models: in the absence of an external bias, bunching proceeds slowly, and
substantially longer simulations are needed for well-developed bunches to form.

Despite these differences, similar patterns emerge in both modeling approaches.
We have identified corresponding parameter combinations across both frameworks
as a first step toward full correspondence. Future work will analyze the
spatiotemporal evolution of these structures; by examining geometric profiles
and conducting rigorous quantitative analysis, we aim to precisely calibrate the
relationship between the discrete and continuous models. Importantly, our goal
was to model the same phenomena—bunching, meandering, and their
coexistence—across both homo- and heteroepitaxial contexts. To achieve this, we
developed specialized meso- and microscale models. Ultimately, the resulting
morphologies from both experiments and models are strikingly similar,
highlighting a clear context-invariance in these instabilities.

\begin{figure}
    \centering
    \includegraphics[
        width=\linewidth,
        height=0.38\textheight,
        keepaspectratio
    ]{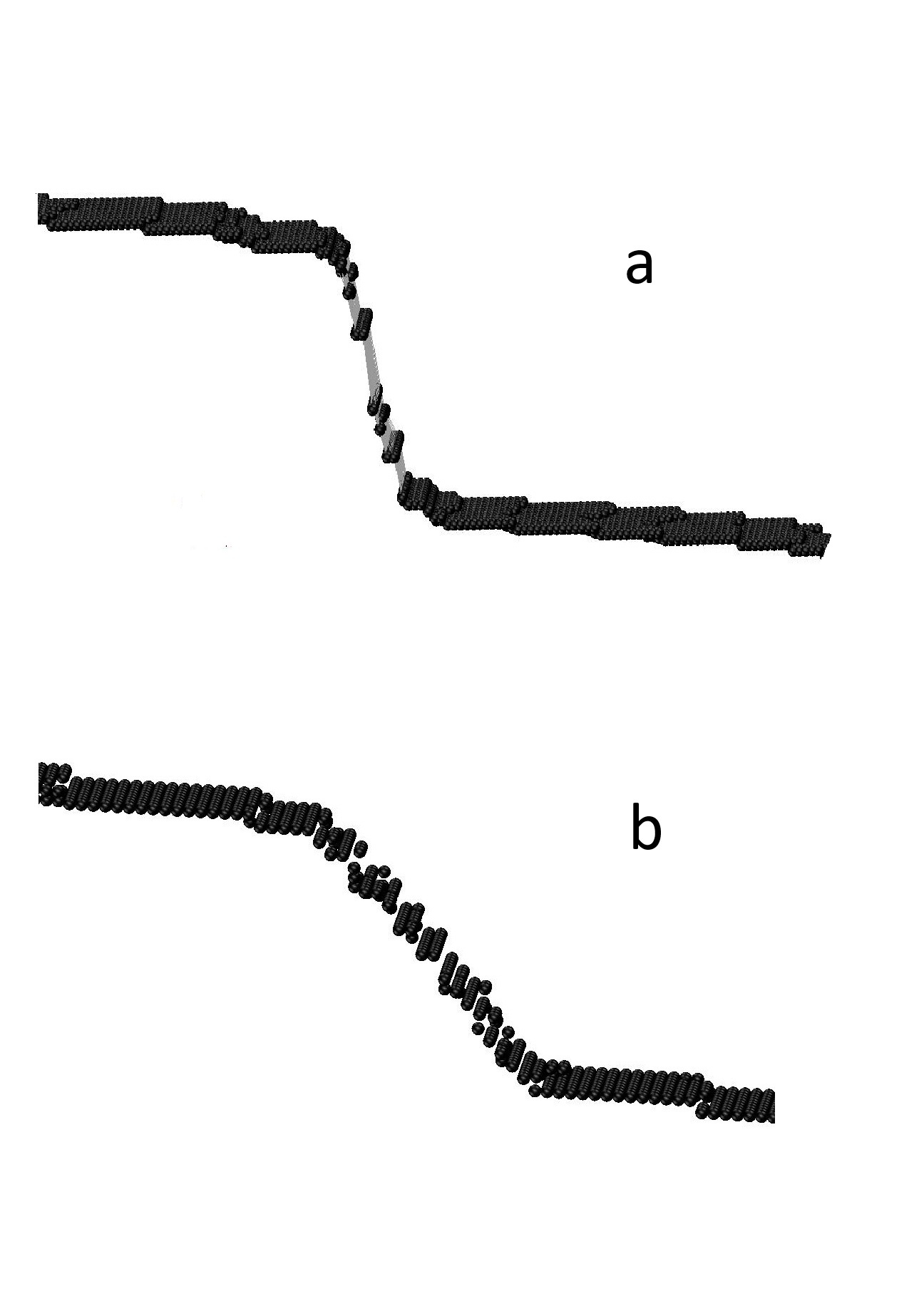}
    \caption{
        Presented as an orthogonal side-view projection of the 3D surface, the
        profiles depict individual surface atoms as points and illustrate the
        bunches obtained for a system of size $2000 \times 10$ with parameters
        $E_S = 3k_BT$, $E_V = 6k_BT$, $c_0 = 0.02$, and $l_0 = 5$. Panels (a)
        and (b) correspond to $n_{DS} = 1$ and $n_{DS} = 50$ diffusion steps,
        respectively, with a simulation time of $2 \times 10^7$.
    }
    \label{fig:profile}
\end{figure}

\begin{figure}
    \centering
    \includegraphics[width=0.9\linewidth]{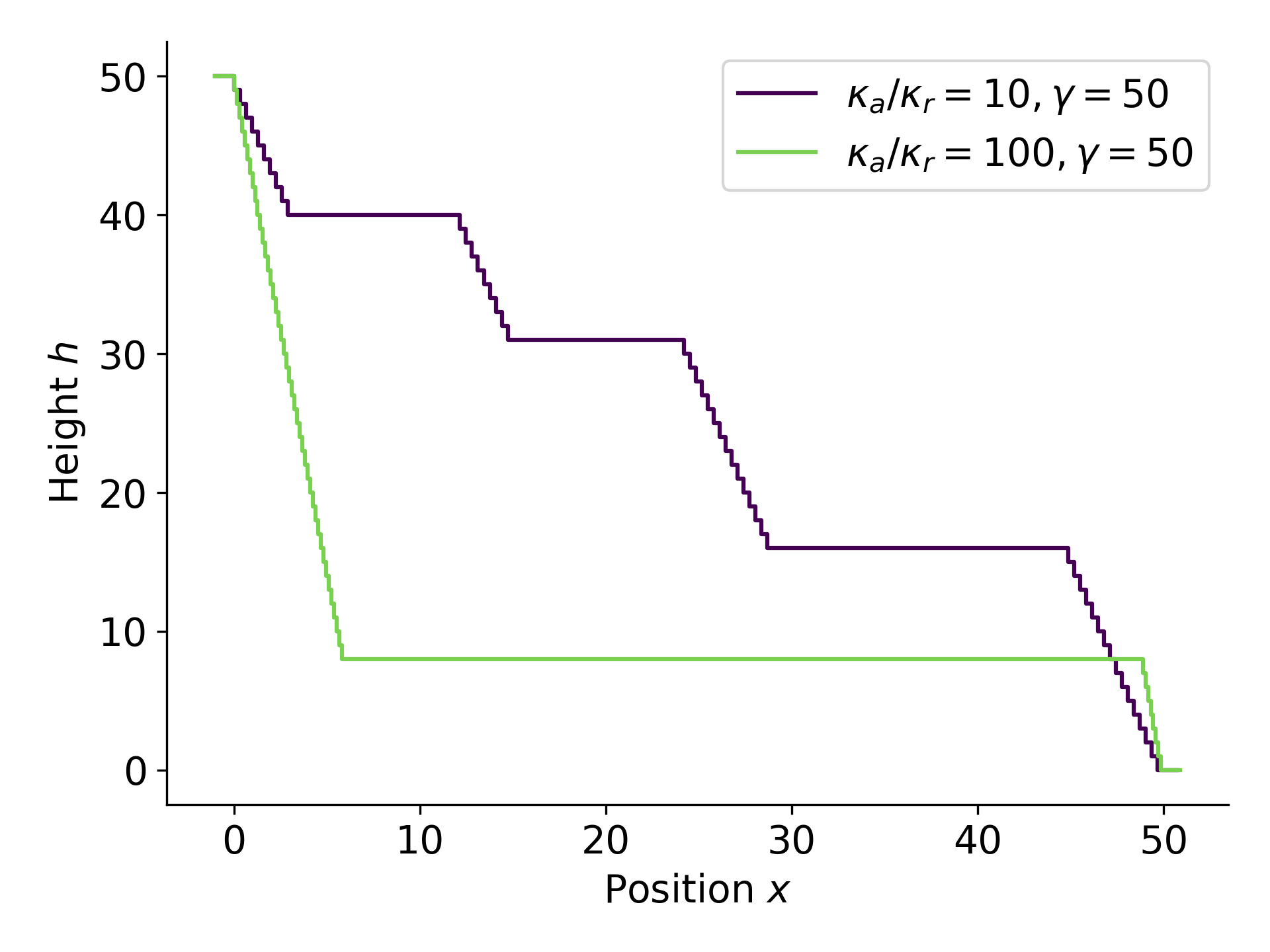}
    \caption{
        Surface profiles obtained for \(N = 50\), \(L = 50\), \(l_0 = 1\),
        \(p = 1\), \(q = 3\) and simulation time \(T = 50\) from
        \autoref{eq:mm2_meandering}. Profiles are taken at exactly the midpoint
        of the domain \(x = L/2 = 25\). Note the bunched ``staircase'', the
        abrupt joining between the bunch and the terrace, and the step-step
        distance independent of the bunch size, all of which are characteristic
        of B1-type bunching. Further, the stronger attraction leads to fewer,
        larger bunches, with lower step-to-step distance inside them.
    }
    \label{fig:mm2_profile}
\end{figure}

Crucially, to identify similar behaviors within the pattern diagram, we limited
our analysis to the parameters $E_S$ and $E_V$, whereas several other parameters
also play an important role. In particular, the number of diffusion steps,
$n_{DS}$, is decisive for the stationary shape of the bunches. To illustrate
this, Fig. \ref{fig:profile} shows two different bunch profiles obtained for
potentials $E_V = 6$ and $E_S = 3$, a case lying in the middle of the bunched
region of the diagram in Fig. \ref{fig:simulations1}. We performed long-term
simulations for $n_{DS} = 1$ and $n_{DS} = 50$ using very thin and very long
systems. As shown in \autoref{fig:profile}, a short diffusion length
(\autoref{fig:profile}a) leads to steep bunches characterized by macrosteps
(highlighted in grey), whereas faster diffusion (\autoref{fig:profile}b) yields
wider bunches with a gentle slope and macrosteps reaching a maximum height of
two. This behavior differs from the profiles of the mesoscopic model in
\autoref{fig:mm2_profile}. In that model, the step-to-step distance within the
bunches depends on the relative difference between attraction and repulsion, but
the profiles otherwise remain close to a sloped straight line and join the
terraces abruptly—contrasting with the smoother, sigmoidal shape present in both
cases of \autoref{fig:profile}.

This indicates that the bunching process operates differently depending on the
diffusion regime. Other parameters, such as the initial terrace width and the
adatom concentration $c_0$, may also influence pattern formation and will be
systematically investigated in future studies.

\section{Conclusions}

In this work we approach the problem of step bunching, step meandering, and
simultaneous bunching and meandering in two different contexts: homo- and
heteroepitaxial growth on vicinal surfaces. We do this through two different
conceptual frameworks - that of a Ginzburg-Landau-type model which coarse-grains
the step edge as a continuous curve and the VicCA model, which operates on the
atomistic level.

The coarse-grained description given by the continuum model ``smooths out'' the
details of attachment-detachment of adatoms to terraces, the steps and kink
positions, resulting in a model that lends itself to both analytical treatment
and the highly-efficient numerical schemes which we implement. This model
directly incorporates the attraction-repulsion between steps through a
Lennard-Jones-type interaction, while ``straightness'' of the step is controlled
by the coarse-grained parameter of step stiffness. This is specifically
constructed for the heteroepitaxial context of logarithmic attraction and
elastic repulsion of steps, and the lack of explicit adatom flux in the model.
With this framework at hand we were able to obtain the full spectrum of
phenomena - bunching, bunching and meandering, and pure meandering and we
construct a morphological diagram for a system with a small number of steps.

Further, we focus on the atomistic VicCA model, which implements the classical
terrace-ledge/step-kink growth rules. The model is aimed at the homoepitaxial
vicinal growth with constant surface flux of diffusing adatoms. We extend the
VicCA by implementing a double-well potential, which defines the jump
probabilities of the diffusing particles. The control parameters of the
double-well VicCA are \(E_V\) - the depth of the potential well in front of the
step and \(E_S\) - the depth of the well on top of the step. We show that this
model setup is then sufficient to describe the above-mentioned spectrum of
instabilities of the regular step-flow. Similarly to the other model, we build a
detailed morphological diagram of the resulting morphologies, when varying
specially chosen combinations of the two potential well energies. Furthermore,
this work demonstrates that complex phenomena such as simultaneous step bunching
and meandering are not model-specific, but rather fundamental growth modes that
persist across different scales of description.

Both diagrams presented in \autoref{fig:3d_diagram} and
\autoref{fig:simulations1} show similar clusters of morphologies, with the
\(\exp(\beta E_V) + \exp( \beta E_S)\) term in the VicCA acting as effective
stiffness and \(\exp( \beta (E_V - E_S))\) being responsible for the
destabilization of the surface. Driven by this numerical matching of qualitative
results, we establish more explicit relations between the two models'
parameters, based on previous results. This matches the accumulated experimental
evidence that simultaneous bunching and meandering is present in both growth
contexts and similar morphologies \textit{should} be expected.

While in this work we focus on the qualitative matching of morphology diagrams,
the relations between meander width in both models pave the way to more
quantitative investigations. For example it is known that the meander width
\(w^2(t) \propto t^n\) where \(n\) is a scaling exponent related to the dominant
mass transport mechanism. Further theoretical understanding of VicCA models such
as an analytical description of the relation of the potential landscape and the
evolving surface would provide understanding which is often not available for
other purely computational models such as kinetic Monte Carlo methods.

Finally, the two models provide a high-performance digital playground,
especially for developing monitoring schemes for the quantification of
real-world phenomena - a critical step for mastering precise surface control for
advanced device manufacturing.

\section*{Acknowledgments}

V.I. and V.T. (BNSF No. KP-06-DO02/1/18.05.2023), H.P. (BNSF No.
KP-06-DO02/2/18.05.2023) and M.A.Ch. and M.A.Z.-K. (NCBR, EIG
CONCERT-JAPAN/9/56/AtLv-AlGaN/2023) are partially financed by the EIG
Concert-Japan project ``Atomic-level control of AlGaN hetero-interfaces for
deep-UV LED (AtLv-AlGaN)'', and express their gratitude to Yoshihiro Kangawa
(PI) from the RIAM at Kyushu University.

Parts of the calculations were done on the HPC facility Nestum
(BG161PO003-1.2.05), HPC resources of the ``National Centre of Excellence
Mechatronics and Clean Technologies'' (Project~№~BG16RFPR002-1.014-0006,
co-funded by the European Union under ``Research Innovation and Digitization for
Smart Transformation'' program 2021-2027) and HPC resources of the GATE
Institute, Bulgaria, via the programme ``Research, Innovation and Digitalisation
for Smart Transformation'' 2021-2027 (PRIDST) (grant agreement no.
BG16RFPR002-1.014-0010-C01). V.I. would like to further thank Boris Kraychev
from GATE Institute, Bulgaria for the support with the computational resources.
V.T. and V.I. thank Sophia Ivanovska from the HPC Hemus for the interest in this
study.

\bibliographystyle{unsrtnat}
\bibliography{bibliography}

\end{document}